\documentclass[aps,pra,twocolumn,floatix,amssymb,amsmath,superscriptaddress]{revtex4-1}
\usepackage{graphicx}
\usepackage{braket}
\usepackage{soul}
\usepackage[usenames,dvipsnames]{color}
\usepackage{epsfig,color}
\usepackage{amsmath,amssymb}
\usepackage{mathptmx}
\usepackage[T1]{fontenc}
\usepackage{verbatim}
\usepackage{float}
\usepackage{units}
\usepackage{bm}
\usepackage{multirow}
\usepackage{booktabs}
\usepackage[colorlinks = true,
            linkcolor = NavyBlue,
            urlcolor  = NavyBlue,
            citecolor = NavyBlue,
            anchorcolor = blue]{hyperref}
\usepackage{epstopdf}
\usepackage[normalem]{ulem}
\usepackage{bbold}
\begin{document}
\title{Including many-body effects into the Wannier-interpolated quadratic photoresponse tensor}
\author{Peio Garcia-Goiricelaya}
\email{peio.garcia@ehu.eus}
\affiliation{Centro de F\'{i}sica de Materiales, Universidad del Pa\'{i}s Vasco UPV/EHU, 20018 San Sebasti\'{a}n, Spain}
\author{Jyoti Krishna}
\affiliation{Centro de F\'{i}sica de Materiales, Universidad del Pa\'{i}s Vasco UPV/EHU, 20018 San Sebasti\'{a}n, Spain}
\author{Julen Iba\~nez-Azpiroz}
\affiliation{Centro de F\'{i}sica de Materiales, Universidad del Pa\'{i}s Vasco UPV/EHU, 20018 San Sebasti\'{a}n, Spain}
\affiliation{Ikerbasque Foundation, 48013 Bilbao, Spain}
\date{\today}
\begin{abstract}
We present a first-principles scheme for incorporating many-body interactions into the unified description of the quadratic optical response to light of noncentrosymmetric crystals.
The proposed method is based on time-dependent current-density response theory and includes the electron-hole attraction \textit{via} a tensorial long-range exchange-correlation kernel, which we calculate using the parameter-free bootstrap  approximation.
By bridging with the Wannier-interpolation of the independent-particle transition matrix elements, the resulting numerical scheme is very general and allows resolving narrow many-body spectral features at low computational cost.
We showcase its potential by inspecting the second-harmonic generation in the benchmark zinc-blende semiconductor GaAs, the layered graphitic semiconductor BC$_{2}$N and the Weyl semimetal TaAs.
Our results show that excitonic effects can give rise to large and sharply localized one- and two-photon resonances that are absent in the independent-particle approximation.
We find overall good agreement with available experimental measurements, capturing the magnitude and peak-structure of the spectrum as well as the  angular dependence at fixed photon energy.
The implementation of the method in Wannier-based code packages can serve as a basis for performing accurate theoretical predictions of quadratic optical properties in a vast pool of materials.
\end{abstract}
\maketitle
\section{INTRODUCTION}
The field of nonlinear optics~\cite{belinicher-jetp82,sturman-book92} has received a considerable push in recent years, thanks in part to advances of  contemporary techniques in designing novel structures such as layered materials and thin films~\cite{Chhowalla2013,Li2014}.
Breakthroughs have come  in various fronts like topology~\cite{doi:10.1126/sciadv.1501524}, with an accute enhancement of the nonlinear light absorption in Weyl semimetals~\cite{Wu2017,Ma2019,Osterhoudt2019} or the prediction of a quantized photoresponse~\cite{deJuan2017}; but also in more applied aspects like the increasing of power-conversion efficiency in ferroelectric insulators~\cite{Spanier2016} or the engineering of new effects for boosting the performance of standard solar cells~\cite{doi:10.1126/science.aan3256}. 

The unified microscopic description of nonlinear optical phenomena is due to Sipe and co-workers~\cite{PhysRevB.48.11705,PhysRevB.52.14636,PhysRevB.61.5337}, who developed a general formalism within the independent-particle approximation for calculating the intrinsic contribution to the second-order optical photoresponse tensors.
This approach accounts for the various quadratic optical processes taking place in semiconductors, including injection and  shift currents~\cite{PhysRevB.23.5590,Fridkin2001,sturman-book92} that originate from physical divergences of the response coefficients.
Building on this scheme, several studies based on density functional theory (DFT) have reported material-specific calculations for various second-order processes;  see Refs.~\cite{PhysRevB.53.10751,PhysRevB.57.3905,PhysRevB.72.045223,PhysRevB.58.7761,PhysRevB.96.115147,Sharma_2004,PhysRevLett.109.116601,Tan2016,PhysRevLett.119.067402,PhysRevResearch.4.L022022} for a small survey.
In addition, recent works have extended the formalism to include metallic terms~\cite{PhysRevResearch.2.012017,PhysRevResearch.3.L042032} and third-order contributions~\cite{PhysRevLett.121.176604,PhysRevLett.126.259701,PhysRevB.100.064301}.
Alternative approaches have also been proposed, \textit{e.g.}, based on the reduced density matrix formalism~\cite{PhysRevB.96.035431}.

While the theory of nonlinear optical photoresponses in the independent-particle approximation has an ample track record, much fewer studies have considered many-body interactions beyond this picture.
Among those, a series of works by Luppi, H\"ubener and Veniard~\cite{hubener1,hubener2,PhysRevB.82.235201,PhysRevB.83.115205} casted the second-order susceptibility within the time-dependent DFT (TDDFT), and provided explicit calculations of excitonic effects on the second-harmonic generation (SHG) spectrum for various materials.
DFT-based SHG spectra influenced by the electron-hole attraction within a Bethe-Salpeter scheme were also reported in 
Refs.~\cite{PhysRevB.65.035205,PhysRevB.71.195209,PhysRevB.96.235206}.
An alternative real-time approach based on the Berry-phase formulation of the dynamical polarization was set forth in Ref.~\cite{PhysRevB.88.235113,PhysRevB.89.081102}.
More recently, quasiparticle and excitonic effects on the shift current 
have been analyzed using the $GW$ plus Bethe-Salpeter equation method~\cite{PhysRevB.101.045104,doi:10.1073/pnas.1906938118}.

The relative scarcity of practical implementations is in part a consequence of the technical difficulties involved.
An important bottleneck concerns the calculation of the independent-particle quadratic response, due to the intricate form of the transition matrix elements that involve derivatives with respect to the crystal momentum $\mathbf{k}$ of Bloch states~\cite{PhysRevB.61.5337}.
Their calculation requires a careful treatment in order to ensure $\mathbf{k}$-space gauge invariance and properly handle 
band degeneracies~\cite{Blount,PhysRevLett.109.116601,esteve-paredes_comprehensive_2023}, 
and brute-force approaches quickly become time-demanding from the computational point of view~\cite{PhysRevLett.119.067402}.
Recently, it has been shown that the so-called ``Wannier interpolation'' procedure can solve the above difficulties~\cite{PhysRevB.97.245143,PhysRevB.96.115147}.
In this approach, the quadratic matrix elements are reformulated in terms of localized Wannier functions, in the same spirit as the Wannier interpolation of the Berry curvature and anomalous Hall conductivity~\cite{PhysRevB.74.195118}.
The method offers a general and efficient way of calculating second-order optical response tensors without band-truncation errors, and can serve as the basis for further developments.

In this work, we incorporate many-body interactions into the Wannier-based scheme by working out an expression for the quadratic optical photoresponse tensor beyond the independent-particle approximation.
Our derivation is based on the time-dependent current-density response theory and formally includes excitonic effects through a tensorial long-range exchange-correlation (xc) kernel.
Explicitly adopting the tensorial character of the response is of central importance, as this allows a natural connection with the formalism of the independent-particle picture in the optical limit and, by extension, with the Wannier-interpolation scheme.
To illustrate the generality and accuracy of our method, we analyze the SHG process in three bulk materials.
In first place, we consider GaAs as a benchmark test.
Secondly, we study BC$_{2}$N, a highly anisotropic graphitic-layered semiconductor that showcases the advantages of the adopted tensorial framework.
Finally, we apply our scheme to the Weyl semimetal TaAs and discuss the 
results in the context of recent optical measurements.

The paper is organized as follows.
In Sec.~\ref{sec:theory} we present the main theoretical scheme. 
We first express the microscopic quadratic conductivity tensor renormalized by many-body interactions, and compare our main tensorial expression with the scalar counterpart of TDDFT~\cite{hubener1,hubener2,PhysRevB.82.235201,PhysRevB.83.115205}.
We then consider the optical limit and specialize to the SHG process, for which we derive new metallic terms.
In order to establish the link to experimental observables, we analyze the connection between the microscopic and macroscopic scales.
Technical details concerning the electronic-structure \textit{ab initio} calculations based on maximally localized Wannier functions and the inclusion of excitonic effects are described in Sec.\ref{sec:methods}.
The computed SHG spectra of GaAs, BC$_{2}$N and TaAs are presented and 
discussed in Sec.~\ref{sec:results_discussion}.
We provide concluding remarks in Sec.~\ref{sec:conclusions},
while several technical subjects are kept for the Appendix.
\section{THEORETICAL FRAMEWORK}\label{sec:theory}
\subsection{Microscopic response tensors and many-body effects}\label{mt}
Our starting point considers the microscopic response of a many-body (MB) system of electrons interacting via the Coulomb potential in a crystal that relates the electric current-density vector $\mathbf{J}(\mathbf{r},t)$ to the powers of an externally applied time-dependent electric field $\mathbf{E}_\mathrm{ext}(\mathbf{r},t)$.
In practice, this amounts to expanding the current-density vector in a power series
\begin{equation}\label{jeqsumnjj}
\mathbf{J}(\mathbf{r},t)=\sum_{j}\mathbf{J}_{j}(\mathbf{r},t),
\end{equation}
with the $j^{\mathrm{th}}$-order contribution defined as
\begin{equation}\label{jneqsigmaprodneext}
 \mathbf{J}_{j}(1)=\int...\int^{1}_{0}\overline{\sigma}_{j}(1,...,j+1)\prod_{j}\mathbf{E}_\mathrm{ext}(j+1)dj+1,
\end{equation}
where we adopted the notation $(\mathbf{r}_{j},t_{j})\equiv(j)$ with $j$ a positive integer.
The quantity $\overline{\sigma}_{j}(1,...,j+1)$ denotes the $j^{\mathrm{th}}$-order MB conductivity tensor, and our main goal consists in finding an expression for the second order, \textit{i.e.}~the $j=2$ contribution.

To do so, let us adopt the standpoint of an electron in an auxiliary Kohn-Sham (KS) system of independent particles, where the \emph{total} electric field that it feels can be written as
\begin{equation}\label{etot}
\mathbf{E}_\mathrm{tot}(\mathbf{r},t)=\mathbf{E}_\mathrm{ext}(\mathbf{r},t)+\mathbf{E}_\mathrm{H}(\mathbf{r},t)+\mathbf{E}_\mathrm{xc}(\mathbf{r},t).
\end{equation}
The Hartree (H) electric field as a function of the current-density vector is given by 
\begin{equation}\label{eh}
 \mathbf{E}_\mathrm{H}(1)=\int^{1}_{0}\overline{K}_\mathrm{H}(1,2)\mathbf{J}(2)d2,
\end{equation}
where $\overline{K}_\mathrm{H}(1,2)$ is the tensorial Hartree kernel.
In turn, the xc electric field up to second order can be written as
\begin{equation}\label{exc}
 \begin{split}
  \mathbf{E}_\mathrm{xc}(1)=&\int^{1}_{0}\overline{K}_{\mathrm{xc},1}(1,2)\mathbf{J}(2)d2\\&+\iint^{1}_{0}\overline{K}_{\mathrm{xc},2}(1,2,3)\mathbf{J}(2)\mathbf{J}(3)d2d3,
 \end{split}
\end{equation}
with $\overline{K}_{\mathrm{xc},1}(1,2)$ and $\overline{K}_{\mathrm{xc},2}(1,2,3)$ the first-order and second-order tensorial xc kernels, respectively.

Within the KS system, the response properties are governed by the so-called KS conductivity tensor, which describes the current-density vector in terms of powers of the total electric field, in such a way that
\begin{equation}\label{jneqsigmaksprodnetot}
 \mathbf{J}_{j}(1)=\int...\int^{1}_{0}\overline{\sigma}^{\mathrm{KS}}_{j}(1,...,j+1)\prod_{j}\mathbf{E}_\mathrm{tot}(j+1)dj+1.
\end{equation}
Thus, the task is to express the MB response tensors up to second order in terms of the necessary KS response coefficients as well as the tensorial Hartree and xc kernels.
\subsubsection{Linear response}\label{tdrt1}
As a warmup, we first review the linear case.
Within the time-dependent current-density functional theory (TDCDFT) \cite{PhysRevLett.77.2037,PhysRevLett.79.4878}, the first-order current-density vector is given by
\begin{subequations}
 \begin{equation}\label{j1ext}
  J^{a}_{1}(1)=\int^{1}_{0}\sum_{b}\sigma_{1}^{ab}(1,2){E_\mathrm{ext}^{b}}(2)d2,
 \end{equation}
 \begin{equation}\label{j1tot}
  J^{a}_{1}(1)=\int^{1}_{0}\sum_{b}\sigma_{1}^{\mathrm{KS},ab}(1,2){E_\mathrm{tot}^{b}}(2)d2
 \end{equation}
\end{subequations}
where $\overline{\sigma}_{1}(1,2)$ and $\overline{\sigma}^{\mathrm{KS}}_{1}(1,2)$ are the first-order conductivity tensors of the MB and the KS system, respectively, defined as
\begin{subequations}
 \begin{equation}\label{sigma1}
  \sigma_{1}^{ab}(1,2)=\frac{\delta J^{a}(1)}{\delta E_\mathrm{ext}^{b}(2)},
 \end{equation}
 \begin{equation}\label{sigma10}
  \sigma_{1}^{\mathrm{KS},ab}(1,2)=\frac{\delta J^{a}(1)}{\delta{E_\mathrm{tot}^{b}}(2)}.
 \end{equation}
\end{subequations}
Henceforth, superscripts refer to Cartesian components.
Applying the chain rule in the definition of the MB conductivity tensor in Eq.~\ref{sigma1} and taking into account the definition of the KS conductivity tensor in Eq.~\ref{sigma10}, the first-order Dyson-like equation relating the MB and KS responses reads
\begin{equation}\label{sigma1sigma10}
  \sigma_{1}^{ab}(1,2)=\int\sum_{c}\sigma_{1}^{\mathrm{KS},ac}(1,3)\epsilon^{-1,cb}(3,2)d3,
\end{equation}
where we have introduced the dielectric tensor $\overline{\epsilon}(1,2)$.
This quantity accounts for the MB electronic screening effects within the crystal and its inverse links the total and external electric fields as
\begin{equation}\label{etot2eext}
 {E_\mathrm{tot}^{a}}(1)=\int^{1}_{0}\varepsilon^{-1,ab}(1,2){E_\mathrm{ext}^{b}}(2)d2.
\end{equation}
Considering the implicit definition of the inverse dielectric tensor in Eq.~\ref{etot2eext} together with the relation between the total and external electric fields in Eq.~\ref{etot}, we apply again the chain rule to obtain
\begin{equation}\label{detotdeext}
 \begin{split}
  \epsilon^{-1,ab}(1,2)=&\delta(1,2)\delta_{ab}\\&+\int\sum_{c}{K_{\mathrm{Hxc},1}^{ac}}(1,3)\sigma_{1}^{cb}(3,2)d3.
 \end{split}
\end{equation}
Above, \mbox{$\overline{K}_{\mathrm{Hxc},1}(1,2)=\overline{K}_\mathrm{H}(1,2)+\overline{K}_{\mathrm{xc},1}(1,2)$} is the grouping of the first-order tensorial Hartree and xc kernels.

For practical purposes, it is useful to express the dielectric tensor in terms of the KS conductivity tensor instead of the MB one.
Such expression is obtained by reproducing the previous chain rule procedure, but this time starting from Eq.~\ref{sigma10}, and reads
\begin{equation}\label{deextdetot}
 \begin{split}
  \epsilon^{ab}(1,2)=&\delta(1,2)\delta_{ab}\\&-\int\sum_{c}{K_{\mathrm{Hxc},1}^{ac}}(1,3)\sigma_{1}^{\mathrm{KS},cb}(3,2)d3.
 \end{split}
\end{equation}
\subsubsection{Quadratic response}\label{tdrt2}
In analogy with the treatment of the first-order response, the second-order current-density vector can be written as
\begin{subequations}
 \begin{equation}\label{j2ext}
  J^{a}_{2}(1)=\iint^{1}_{0}\sum_{bc}\sigma_{2}^{abc}(1,2,3){E_\mathrm{ext}^{b}}(2){E_\mathrm{ext}^{c}}(3)d2d3,
 \end{equation}
 \begin{equation}\label{j2tot}
  J^{a}_{2}(1)=\iint^{1}_{0}\sum_{bc}\sigma_{2}^{\mathrm{KS},abc}(1,2,3){E_\mathrm{tot}^{b}}(2){E_\mathrm{tot}^{c}}(3)d2d3,
 \end{equation}
\end{subequations}
where $\overline{\sigma}_{2}(1,2,3)$ and $\overline{\sigma}^\mathrm{KS}_{2}(1,2,3)$ are the second-order conductivity tensors of the MB and the KS systems, respectively, defined as
\begin{subequations}
 \begin{equation}\label{sigma2}
  \sigma_{2}^{abc}(1,2,3)=\frac{\delta^{2} J^{a}(1)}{\delta {E_\mathrm{ext}^{b}}(2)\delta{E_\mathrm{ext}^{c}}(3)},
 \end{equation}
 \begin{equation}\label{sigma20}
  \sigma_{2}^{\mathrm{KS},abc}(1,2,3)=\frac{\delta^{2} J^{a}(1)}{\delta {E_\mathrm{tot}^{b}}(2)\delta{E_\mathrm{tot}^{c}}(3)}.
 \end{equation}
\end{subequations}
By sistematically applying the chain rule in the definition of the MB conductivity tensor in Eq.~\ref{sigma2}, a procedure that is outlined in Appendix~\ref{appxjjj}, one derives the desired second-order Dyson-like equation relating the MB and KS responses 
\begin{widetext}
\begin{equation}\label{sigma2de}
 \begin{split}
  \sigma_{2}^{abc}(1,2,3)=&\iiint\sum_{def}\varepsilon^{-1,ad}(1,4)\sigma_{2}^{\mathrm{KS},def}(4,5,6)\varepsilon^{-1,eb}(5,2)\varepsilon^{-1,fc}(6,3)d4d5d6\\
  &+\iiint\sum_{def}\sigma_{1}^{ad}(1,4)K_{\mathrm{xc},2}^{def}(4,5,6)\sigma_{1}^{eb}(5,2)\sigma_{1}^{fc}(6,3)d4d5d6.
 \end{split}
\end{equation}
\end{widetext}
The above equation can be regarded as the tensorial generalization of the expression for the second-order scalar density response function obtained in TDDFT (see Eq.~180 in Ref.~\cite{Gross1996} or Eq.~13 in Ref.~\cite{hubener1}).
Dealing with the response in the form of a tensorial quantity allows a natural connection with the description of the optical KS response, as we show below.
\subsection{Optical limit}
\label{sec:optical}
To proceed further, one needs explicit expressions for the KS response.
This task can be greatly simplified by considering the optical and long-wavelength limit, which assumes that the external electric field remains constant in the length-scale of the crystal's unit cell~\cite{Ehrenreich}.
Within this approach, it is convenient to adopt the formalism of Sipe and co-workers~\cite{PhysRevB.48.11705,PhysRevB.52.14636,PhysRevB.61.5337}, where the current-density vector operator is split into its interband (ter) and intraband (tra) parts at any $j^{\mathrm{th}}$ order as
\begin{equation}\label{JeqdPter+Jtra}
 \mathbf{J}_{j}(t)=\frac{d\mathbf{P}_{\mathrm{ter},j}(t)}{dt}+\mathbf{J}_{\mathrm{tra},j}(t),
\end{equation}
where the interband polarization- and intraband current-density vectors are respectively expressed in terms of the charge-density matrix elements $\rho_{j,mn}(t)$ as
\begin{subequations}
 \begin{equation}\label{Pter}
  P^{a}_{\mathrm{ter},j}(t)=\frac{e}{V}\sum_{\mathbf{k}mn}r^{a}_{nm}\rho_{j,mn}(t),
 \end{equation}
 \begin{equation}\label{Jtra}
  \begin{split}
   J&^{a}_{\mathrm{tra},j}(t)=\\&\frac{e}{V}\sum_{\mathbf{k}mn}\left[v^{a}_{nm}\delta_{nm}-\sum_{b}\left(r^{a;b}_{nm}+\delta_{nm}\epsilon_{cab}\Omega^{c}_{n}\right)\right]\rho_{j,mn}(t).
  \end{split}
 \end{equation}
\end{subequations}
Above, $V$ denotes the volume of the crystal, while $n$ and $m$ are band indices.
The transition matrix elements involve several quantities;
\mbox{${r}^{a}_{nm}=(1-\delta_{nm}){\xi}^{a}_{nm}$} and \mbox{$r^{a;b}_{nm}=\partial r^{a}_{nm}/\partial k^{b}-i(\xi^{b}_{nn}-\xi^{b}_{mm})r^{a}_{nm}$} are the interband dipole and its generalized derivative, respectively;
\mbox{$\xi^{a}_{nm}=i\braket{u_{n}|\partial/\partial k^{a}|u_{m}}$} and \mbox{$\epsilon_{cab}\Omega^{c}_{n}=\partial\xi^{b}_{nn}/\partial k^{a}-\partial\xi^{a}_{nn}/\partial k^{b}$} stand for the Berry connection and curvature, respectively, with $\ket{u_{n}}$ the crystal-periodic part of the Bloch function;
finally, \mbox{$v^{a}_{nm}=\braket{u_{n}|\partial\hat{H}/\partial k^{a}|u_{m}}$} denotes the velocity matrix element.
We kept the dependence on the crystal wave vector $\mathbf{k}$ implicit for all these quantities.

Based on the dynamical equation of the charge-density operator within the Schr\"odinger picture, one can solve for $\rho_{j,mn}(t)$ employing an iterative scheme at the desired order in the electric field and compute the associated response tensors~\cite{PhysRevB.61.5337}.
In frequency domain, the first-order interband polarization- and intraband current-density vectors can be respectively expressed as
\begin{subequations}
 \begin{equation}\label{P1ter}
  P^{a}_{\mathrm{ter},1}(\omega)=\sum_{b}\alpha^{\mathrm{KS},ab}_{\mathrm{ter},1}(\omega)E^{b}_{\mathrm{tot}}(\omega),
 \end{equation}
 \begin{equation}\label{J1tra}
  J^{a}_{\mathrm{tra},1}(\omega)=\sum_{b}\sigma^{\mathrm{KS},ab}_{\mathrm{tra},1}(\omega)E^{b}_{\mathrm{tot}}(\omega),
 \end{equation}
\end{subequations}
and similarly for second order
\begin{subequations}
 \begin{equation}\label{P2ter}
  P^{a}_{\mathrm{ter},2}(\omega_{12})=\sum_{bc}\alpha^{\mathrm{KS},abc}_{\mathrm{ter},2}(\omega_{1},\omega_{2})E^{c}_{\mathrm{tot}}(\omega_{1})E^{b}_{\mathrm{tot}}(\omega_{2}),
 \end{equation}
 \begin{equation}\label{J2tra}
  J^{a}_{\mathrm{tra},2}(\omega_{12})=\sum_{bc}\sigma^{\mathrm{KS},abc}_{\mathrm{tra},2}(\omega_{1},\omega_{2})E^{b}_{\mathrm{tot}}(\omega_{1})E^{c}_{\mathrm{tot}}(\omega_{2}),
 \end{equation}
\end{subequations}
with $\omega_{12}=\omega_{1}+\omega_{2}$.
In Eqs.~\ref{P1ter} and \ref{J1tra}, $\overline{\alpha}^{\mathrm{KS}}_{\mathrm{ter},1}(\omega)$ and $\overline{\sigma}^{\mathrm{KS}}_{\mathrm{tra},1}(\omega)$ are the first-order optical KS interband polarizability and intraband conductivity tensors, respectively, while in Eqs.~\ref{P2ter} and \ref{J2tra}, $\overline{\alpha}^{\mathrm{KS}}_{\mathrm{ter},2}(\omega_{1},\omega_{2})$ and $\overline{\sigma}^{\mathrm{KS}}_{\mathrm{tra},2}(\omega_{1},\omega_{2})$ are their second-order counterparts, respectively.
Following Eq.~\ref{JeqdPter+Jtra}, the full optical KS conductivity tensors at first and second order are respectively given by
\begin{equation}
 \overline{\sigma}_{1}^{\mathrm{KS}}(\omega)=-i\omega{\overline{\alpha}_{\mathrm{ter},1}^{\mathrm{KS}}}(\omega)+{\overline{\sigma}_{\mathrm{tra},1}^{\mathrm{KS}}}(\omega),
\end{equation}
and
\begin{equation}
 \overline{\sigma}_{2}^{\mathrm{KS}}(\omega_{1},\omega_{2})=-i\omega_{12}{\overline{\alpha}_{\mathrm{ter},2}^{\mathrm{KS}}}(\omega_{1},\omega_{2})+{\overline{\sigma}_{\mathrm{tra},2}^{\mathrm{KS}}}(\omega_{1},\omega_{2}).
\end{equation}

The expressions for the optical KS interband polarizability and intraband conductivity tensors are well established at first order~\cite{doi:10.1143/JPSJ.12.570,doi:10.1143/JPSJ.12.1203}, as well as at second order in the case of semiconductors~\cite{PhysRevB.48.11705,PhysRevB.52.14636,PhysRevB.61.5337}.
In the case of metals and semimetals extra terms appear due to the presence of a Fermi surface.
Recent works~\cite{PhysRevResearch.2.012017,PhysRevResearch.3.L042032} have derived and thoroughly discussed the metallic terms of $\overline{\sigma}^{\mathrm{KS}}_{\mathrm{tra},2}(\omega_{1},\omega_{2})$, paying special attention to the direct-current contribution.
As for $\overline{\alpha}^{\mathrm{KS}}_{\mathrm{ter},2}(\omega_{1},\omega_{2})$, its metallic terms have not been previously derived to the best of our knowledge.
In Appendix \ref{app:general} we provide the general expressions of all optical KS response tensors up to second order valid for any kind of material.
\hspace{1mm}\\
\subsubsection{Second harmonic generation}
In the remaining of this work, for conciseness we specialize in the calculation of a particular quadratic optical response, namely the second harmonic generation.
The SHG process considers two initial photons with same frequency which are combined to generate a final photon with twice the initial frequency, maintaining the coherence of the excitation.
By setting $\omega_{1}=\omega_{2}\equiv\omega$ in Eqs.~\ref{alphater20} and \ref{sigmatra20} of Appendix \ref{app:general}, the SHG KS interband polarizability intraband conductivity tensors are respectively given by
\begin{widetext}
\begin{subequations}
 \begin{equation}\label{alphatershg}
  \begin{split}
   \alpha^{\mathrm{KS},abc}_{\mathrm{ter},2}&(\omega,\omega)=\\&\frac{e^{3}}{2\hbar^{2}V}\Bigg(\sum_{\mathbf{k}mnl}\frac{r^{a}_{nm}\left(r^{b}_{ml}r^{c}_{ln}+r^{c}_{ml}r^{b}_{ln}\right)}{\omega_{ln}-\omega_{ml}}\bigg(\frac{2f_{nm}}{\omega_{mn}-2\tilde{\omega}}-\frac{f_{nl}}{\omega_{ln}-\tilde{\omega}}-\frac{f_{ml}}{\omega_{ml}-\tilde{\omega}}\bigg)+i\sum_{\mathbf{k}mn}\Bigg\{f_{nm}\Bigg[\frac{2r^{a}_{nm}\left(r^{b;c}_{mn}+r^{c;b}_{mn}\right)}{\omega_{mn}\left(\omega_{mn}-2\tilde{\omega}\right)}+\\
   &\frac{r^{a;b}_{nm}r^{c}_{mn}+r^{a;c}_{nm}r^{b}_{mn}}{\omega_{mn}\left(\omega_{mn}-\tilde{\omega}\right)}+\frac{r^{a}_{nm}\left(r^{b}_{mn}\Lambda^{c}_{mn}+r^{c}_{mn}\Lambda^{b}_{mn}\right)}{\omega^{2}_{mn}}\bigg(\frac{1}{\omega_{mn}-\tilde{\omega}}-\frac{4}{\omega_{mn}-2\tilde{\omega}}\bigg)\Bigg]-\frac{r^{a}_{nm}\left(f_{nm;b}r^{c}_{mn}+f_{nm;c}r^{b}_{mn}\right)}{\omega_{mn}\omega}\Bigg\}\Bigg),
  \end{split}
 \end{equation} 
 \begin{equation}\label{sigmatrashg}
  \begin{split}
   \sigma^{\mathrm{KS},abc}_{\mathrm{tra},2}&(\omega,\omega)=\\&\frac{e^{3}}{2\hbar^{2}V}\Bigg\{-\sum_{\mathbf{k}mn}f_{nm}\Bigg[\frac{r^{c;a}_{nm}r^{b}_{mn}+r^{b;a}_{nm}r^{c}_{mn}}{\omega_{mn}-\tilde{\omega}}+\frac{\Lambda^{a}_{nm}\left(r^{b}_{mn}r^{c}_{nm}+r^{c}_{mn}r^{b}_{nm}\right)}{2\omega\left(\omega_{mn}-\tilde{\omega}\right)}\Bigg]+\sum_{\mathbf{k}n}\left[i\frac{\left(f_{n;b}\epsilon_{dac}+f_{n;c}\epsilon_{dab}\right)\Omega^{d}_{n}}{\omega}-\frac{v^{a}_{n}f_{n;bc}}{\omega^{2}}\right]\Bigg\}.
  \end{split}
 \end{equation}
\end{subequations}
%
Above, \mbox{$\omega_{mn}=\omega_{m}-\omega_{n}$} and \mbox{$f_{nm}=f_{n}-f_{m}$}, with \mbox{$\hbar\omega_{n}$} and \mbox{$f_{n}=f(\hbar\omega_{n})$} the eigenvalue and occupation factor of the eigenstate $\ket{\mathbf{k}n}$, respectively, while \mbox{$\Lambda^{a}_{nm}=v^{a}_{n}-v^{a}_{m}$} and \mbox{$\tilde{\omega}\equiv\omega+i\eta/\hbar$}, with $\eta$ a 
positive real infinitesimal.
The terms including the derivatives \mbox{$f_{n;a}=\partial f_{n}/\partial k^{a}$} and \mbox{$f_{n;ab}=\partial^{2}f_{n}/\partial k^{a}\partial k^{b}$} correspond to the metallic contribution.
With explicit expressions for the SHG KS response coefficients at hand, we can now calculate the SHG MB conductivity tensor from Eq.~\ref{sigma2de} as
%
\begin{equation}\label{sigma2deopt}
 \begin{split}
  \sigma_{2}^{abc}(\omega,\omega)=\sum_{def}\varepsilon^{-1,ad}(2\omega)\sigma_{2}^{\mathrm{KS},def}(\omega,\omega)\varepsilon^{-1,eb}(\omega)\varepsilon^{-1,fc}(\omega)+\sum_{def}\sigma_{1}^{ad}(2\omega)K_{\mathrm{xc},2}^{def}(\omega,\omega)\sigma_{1}^{eb}(\omega)\sigma_{1}^{fc}(\omega),
 \end{split}
\end{equation}
\end{widetext}
where the optical dielectric and MB conductivity tensors satisfy respectively (see Eqs.~\ref{deextdetot} and \ref{sigma1sigma10})
\begin{equation}\label{eq:eps_m}
 \begin{split}
  \epsilon^{ab}(\omega)=\delta_{ab}-\sum_{c}K_{\mathrm{Hxc},1}^{ac}(\omega)\sigma_{1}^{\mathrm{KS},cb}(\omega),
 \end{split}
\end{equation}
and
\begin{equation}\label{eq:sigma1}
 \sigma_{1}^{ab}(\omega)=\sum_{c}\sigma_{1}^{\mathrm{KS},ac}(\omega)\epsilon^{-1,cb}(\omega).
\end{equation}

Let us inspect Eq.~\ref{sigma2deopt} in some detail.
MB interactions come in two different ways;
on the one hand, through the inverse dielectric tensor that includes screening effects,
and on the other hand, through the second-order tensorial xc kernel $\overline{K}_{\mathrm{xc},2}(\omega,\omega)$.
Due to hierarchy arguments we expect the former to be dominant.
Focusing on the first piece on the right-hand side (r.h.s.)~of Eq.~\ref{sigma2deopt}, the response at frequency $\omega$ is affected by the screening at that and twice that frequency.
This can lead to double-frequency many-body resonances in the SHG spectrum, 
as we  will show in more detail when analyzing our numerical results in 
Sec.~\ref{sec:results_discussion}.
Note that, in the case of isotropic media, Eq.~\ref{sigma2deopt} is equivalent to Eq.~37 of Ref.~\cite{PhysRevB.82.235201}.
\subsubsection{From the microscopic response to the macroscopic response}\label{sec:macroscopic}
As the final step, we consider the connection between the previous microscopic coefficients and their macroscopic counterparts, which are ultimately the quantities measured in experiment.
The macroscopic response to light is described by Maxwell's equations and can be accessed by performing a macroscopic average of the microscopic response tensors over regions in space that are large in comparison with the crystal unit cell, but small compared to the wavelength of the external perturbation~\cite{Ehrenreich}.
In this work we adopt the formulation of Del Sole and Fiorino~\cite{PhysRevB.29.4631} for relating the macroscopic and microscopic scales; the detailed derivation is outlined in Appendix~\ref{appendix:micro-macro}.
Here we focus on the SHG process; by setting $\omega_{1}=\omega_{2}\equiv\omega$ in Eq.~\ref{sigma2M}, the macroscopic SHG photoconductivity tensor is calculated from its microscopic counterpart as
\begin{equation}\label{micro2macroshg}
 \begin{split}
  \sigma_{\mathrm{M},2}^{abc}(\omega,\omega)=\epsilon_{\mathrm{M}}^{aa}(2\omega)\sigma_{2}^{abc}(\omega,\omega)&\epsilon_{\mathrm{M}}^{bb}(\omega)\epsilon_{\mathrm{M}}^{cc}(\omega)
 \end{split},
\end{equation}
where the macroscopic optical dielectric tensor is given in terms of the microscopic optical conductivity by
\begin{equation}\label{eq:epsM}
 \epsilon^{aa}_{\mathrm{M}}(\omega)=\left[\mathbb{1}-i\frac{4\pi}{\omega}\overline{\sigma}_{1}(\omega)\right]^{-1,aa}.
\end{equation}
\section{TECHNICAL DETAILS}\label{sec:computational_details}
\label{sec:methods}
In this section we describe in detail the steps followed in the calculations for three bulk materials: the semiconductor GaAs, the semiconductor BC$_{2}$N and the semimetal TaAs.
\subsection{DFT calculations}
In a first step, we performed DFT self-consistent calculations using the \textsc{Quantum Espresso} code package~\cite{Giannozzi_2009,Giannozzi_2017}.
The interaction between valence electrons and atomic cores was modeled by means of projector-augmented-wave pseudopotentials~\cite{PhysRevB.50.17953} with scalar relativistic corrections for GaAs and BC$_{2}$N  and fully relativistic corrections for TaAs.
The pseudopotentials were taken from the \textsc{Quantum Espresso} website and generated using the Perdew-Burke-Ernzerhof generalized gradient approximation for thexc energy functional~\cite{PhysRevLett.77.3865}.
For GaAs, we considered the zinc-blende crystal structure together with the experimental value of the lattice parameter, \textit{i.e.}~$a=10.68~a_{0}$~\cite{ashcroft1976solid}.
We performed DFT calculations using a $8\times8\times8$ $k$-point mesh in combination with fixed occupation values and a plane-wave basis set with a cut-off energy of $60~\mathrm{Ry}$.
For BC$_{2}$N, we considered the graphitic-layered A2 crystal structure, which is the most stable noncentrosymmetric bulk structure, with orthorhombic space group $Pmm2$ (No.~25) following the theoretical structural parameters of Ref.~\cite{PhysRevB.73.193304}.
We performed DFT calculations using a $10\times10\times10$ $k$-point mesh in combination with fixed occupation values and a plane-wave basis set with a cut-off energy of $70~\mathrm{Ry}$.
Finally, for TaAs, we considered its ground-state body-centered-tetragonal crystal structure with nonsymmorphic space group $I41md$ (No.~109) following the experimental structural parameters of Ref.~\cite{acta.chem.scand.19-0095}.
We performed non-collinear spin-DFT calculations using a $8\times8\times8$ $k$-point mesh in combination with occupation values calculated by means of the optimized tetrahedron method~\cite{PhysRevB.89.094515} and a plane-wave basis set with a cut-off energy of $60~\mathrm{Ry}$.
\subsection{Wannier interpolation}
\begin{figure}
 \includegraphics[width=\linewidth]{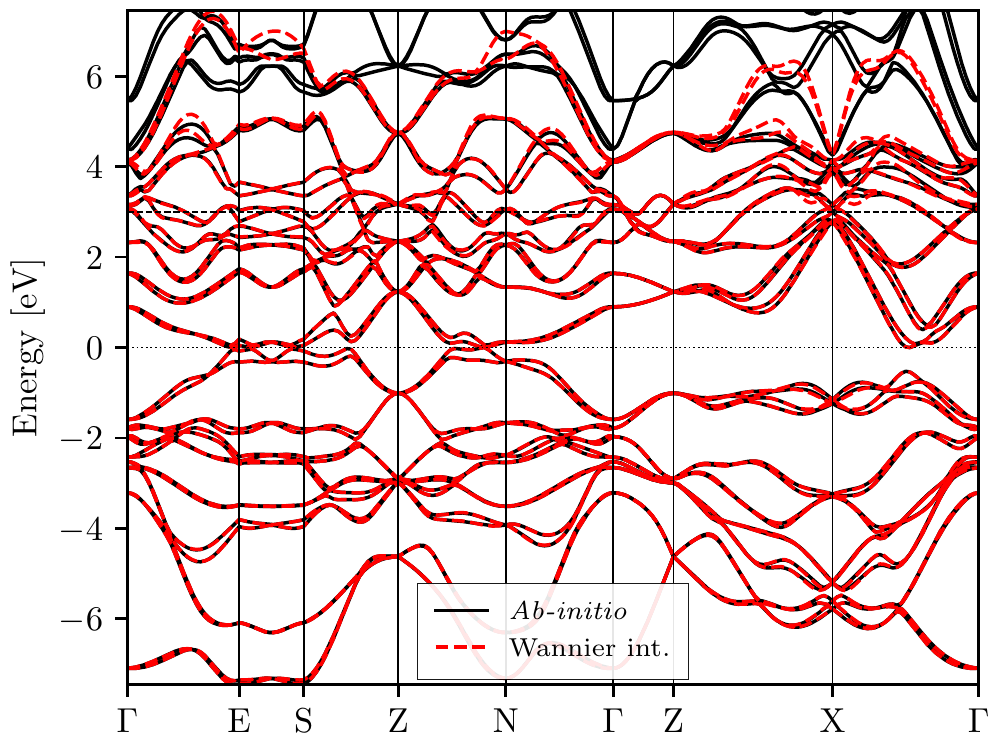}
 \caption{DFT and Wannier-interpolated energy bands of TaAs. The horizontal dashed line at $3~\mathrm{eV}$ denotes the upper limit of the inner energy window used in the disentanglement step of the Wannier construction procedure.}
 \label{fig:bands}
\end{figure}

In a postprocessing step, we constructed maximally localized Wannier functions (MLWF) using the \textsc{Wannier90} code package~\cite{Pizzi_2020}.
For GaAs, starting from a set of 15 spin-degenerate bands, we 
constructed 11 disentangled MLWFs spanning the 4 high-energy valence bands and the 7 low-energy  conduction bands using two $s$ and one $p$ trial orbitals centered on all atoms, as well as one $s$ trial orbital halfway between the two atoms.
For BC$_{2}$N, starting from a set of 38 spin-degenerate bands, we constructed 8 disentangled MLWF spanning the 4 high-energy valence bands and the 4 low-energy conduction bands using $p_{z}$ trial orbitals centered on all atoms.
Finally, for TaAs, starting from a set of 48 spin-polarized bands, we constructed 32 disentangled MLWFs spanning the 16 high-energy valence bands and the 16 low-energy conduction bands using $p$ and $d$ trial orbitals centered on all As and Ta atoms, respectively.
In all cases, the agreement between DFT and Wannier-interpolated bands is in excellent agreement inside the chosen inner energy window~\cite{PhysRevB.65.035109}, as we illustrate in Fig.~\ref{fig:bands} for the case of TaAs.

Having converged the Wannier basis, we then computed the linear and quadratic optical KS response tensors (see Eqs.~\ref{alphater10}-\ref{sigmatra10} and Eqs.~\ref{alphatershg}-\ref{sigmatrashg}, respectively) using Wannier interpolation.
To that end, we used the schemes described in Ref.~\cite{PhysRevB.74.195118} for the calculation of interband dipole matrix elements and Berry curvatures, Ref.~\cite{PhysRevB.97.245143} for the calculation of generalized derivatives of the dipole matrix and Ref.~\cite{PhysRevB.75.195121} for the calculation of velocity matrix elements.
Following the procedure of Refs.~\cite{PhysRevB.97.245143,PhysRevB.103.247101}, in Eqs.~\ref{alphatershg}-\ref{sigmatrashg} we regularized the energy denominators of the three-band term and $r^{a;b}_{mn}$ involving intermediate states by means of an auxiliary parameter $\eta_{\mathrm{r}}$.
Alongside, the derivatives of the occupation factors were computed by replacing
\begin{equation}
 f_{n;a}\to \dfrac{df}{d\omega_n}v^{a}_{n},
\end{equation}
for the first-order derivative and
\begin{equation}
 f_{n;ab}\to \dfrac{d^{2}f}{d\omega^{2}_n}v^{a}_{n}v^{b}_{n}+
 \dfrac{df}{d\omega_n}\omega_{n;ab},
\end{equation}
for the second-order derivative, where $\omega_{n;ab}$ denotes the inverse effective mass tensor~\cite{PhysRevB.75.195121}.
We considered Gaussian distributions for the derivatives of the occupation factors.

In order to obtain well-converged optical spectra, we used dense $k$-point interpolation meshes of $250\times250\times250$ for GaAs, $200\times200\times200$ for BC$_{2}$N, and $300\times300\times300$ for TaAs.
With respect to the imaginary part of the complex energy $\hbar\tilde{\omega}$ (see Eqs.~\ref{alphatershg} and \ref{sigmatrashg}), we set $\eta=0.1~\mathrm{eV}$ in the case of GaAs and TaAs, consistent with carrier scattering lifetimes $(\sim10~\mathrm{fs})$ near the Fermi level observed in both GaAs~\cite{doi:10.1073/pnas.1419446112} and TaAs~\cite{PhysRevResearch.2.013073}, while for BC$_{2}$N we employed an adaptative scheme~\cite{PhysRevB.75.195121}.
Regarding the auxiliary parameter for regularizing energy denominators, we chose $\eta_{r}=0.04~\mathrm{eV}$ for both GaAs and BC$_{2}$N, following Refs.~\cite{PhysRevB.97.245143} and \cite{10.21468/SciPostPhys.12.2.070}, respectively.
In the case of TaAs, we set \mbox{$\eta_{r}=0.1~\mathrm{meV}$}, in order to properly capture the contribution of Weyl points.
The occupation factors and their derivatives are evaluated at zero temperature $(T=0~\mathrm{K})$ for the semiconductors GaAs and BC$_{2}$N and at room temperature $(T=300~\mathrm{K})$ for TaAs.
\subsection{Long-range contribution to the tensorial xc kernel}
Within the long-wavelength and optical limit, the general expressions for the tensorial Hartree and xc kernels (see Eqs.~\ref{eh} and \ref{exc}) simplify.
The Hartree term reduces to a diagonal and isotropic tensor
\begin{equation}\label{optkh}
 K^{ab}_{\mathrm{H}}(\omega)=-i\frac{4\pi\delta_{ab}}{\omega},
\end{equation}
As for the xc term, it takes the form of a long-range contribution (LRC) with a screened Coulomb-like potential, as first discussed in Ref.~\cite{PhysRevB.54.8540}.
In our work, we took into account quasiparticle self-energy effects by means of a scissors operator, while we incorporated electron-hole interactions assuming a static tensorial LRC xc kernel based on an attractive Coulomb-like potential.
With these assumptions, the first-order tensorial xc kernel  simplifies to (see Appendix~\ref{app:xc})
\begin{equation}\label{lrckxc1}
 K^{ab}_{\mathrm{xc},1}(\omega)=i\frac{\alpha_{\mathrm{LRC}}^{a}\delta_{ab}}{\omega},
\end{equation}
which is a diagonal but generally anisotropic $3\times3$ matrix composed of three independent, positive-definite and frequency-independent coefficients $\alpha_{\mathrm{LRC}}^{a}$.
We note that $\alpha^{a}_{\mathrm{LRC}}$ in Eq.~\ref{lrckxc1} is the tensorial generalization of the scalar $\alpha$-parameter of LRC xc kernels used in TDDFT~\cite{PhysRevLett.88.066404,RevModPhys.74.601,PhysRevB.69.155112}.
In our implementation, we calculated these coefficients by means of the so-called self-consistent bootstrap (BO) approximation~\cite{PhysRevLett.107.186401} along each principal axis of the material
(see Appendix~\ref{app:xc} for details).
While this approximation might underestimate excitonic effects in large band-gap insulators~\cite{PhysRevLett.114.146402}, we have verified that our results on semiconducting GaAs and BC$_{2}$N are practically unchanged when using an alternative one-shot RPA-bootstrap (RBO) approximation proposed in Refs.~\cite{PhysRevLett.114.146402,PhysRevLett.115.137402}.

Finally, in our calculations we discarded the effect of the second-order tensorial xc kernel $\overline{K}_{\mathrm{xc},2}(1,2,3)$ entering Eq.~\ref{exc}, given that its approximate expression is generally unknown and its effects are expected to be minor (of the order of crystal local-field effects~\cite{PhysRevB.54.8540})
in comparison to the first-order contribution~\cite{hubener1,hubener2,PhysRevB.82.235201}.

To sum up, in practice, we calculated the microscopic SHG MB conductivity tensor by means of
\begin{equation}\label{ks2mbmicroshg}
 \begin{split}
  \sigma_{2}^{abc}&(\omega,\omega)=\\&\sum_{def}\varepsilon^{-1,da}(2\omega)\sigma_{2}^{\mathrm{KS},def}(\omega,\omega)\varepsilon^{-1,eb}(\omega)\varepsilon^{-1,fc}(\omega),
 \end{split}
\end{equation}
where the microscopic optial dielectric tensor is given by
\begin{equation}\label{eq:epsmbs}
 \epsilon^{ab}(\omega)=\delta_{ab}+i\frac{4\pi-\alpha_{\mathrm{LRC}}^{a}}{\omega}\sigma_{1}^{\mathrm{KS},ab}(\omega).
\end{equation}
The last step involves calculating the macroscopic SHG photoconductivity tensor from its microscopic counterpart by means of Eq.~\ref{micro2macroshg}.
\section{RESULTS}\label{sec:results_discussion}
In this section we present our numerical results of the macroscopic SHG photoresponse.
To facilitate comparison with existing literature, we will partly describe our results in terms of the photosusceptibility, whose connection to the photoconductivity used in our derivations  of Sec.~\ref{sec:theory} is provided in Appendix~\ref{appendix:micro-macro}.
In the materials analyzed in this work, the optical dielectric tensor is
diagonal due to symmetry arguments~\cite{boyd2008nonlinear}.
It then follows that the relation between the macroscopic SHG MB and KS photosusceptibilities simplifies to
\begin{equation}\label{ksmicro2mbmacro_abc}
 \chi^{abc}_{2}(\omega,\omega)=\beta^{abc}(\omega)\chi^{\mathrm{KS},abc}_{2}(\omega,\omega),
\end{equation}
with
\begin{equation}\label{eq:beta}
 \begin{split}
  \beta^{abc}(\omega)=&\varepsilon^{aa}_{\mathrm{M}}(2\omega)\varepsilon^{-1,aa}(2\omega)\times\\&\varepsilon^{bb}_{\mathrm{M}}(\omega)\varepsilon^{-1,bb}(\omega)\varepsilon^{cc}_{\mathrm{M}}(\omega){\varepsilon^{-1}}^{cc}(\omega).
 \end{split}
\end{equation}
the \emph{enhancement factor}, a quantity that will be useful when discussing the impact of MB corrections in our results.
\subsection{GaAs}\label{subsec:gaas}
The first SHG measurements in GaAs date back to the 1960's~\cite{PhysRevLett.10.474,doi:10.1063/1.1713080}, and it has become the standard material for benchmarking theoretical SHG calculations.
Initial works were based on empirical pseudopotentials~\cite{PhysRevB.12.2325} and tight-binding models~\cite{PhysRevB.36.9708}.
More recently, several first principles studies have been reported~\cite{PhysRevB.45.8738,PhysRevB.53.10751,PhysRevB.57.3905,PhysRevB.72.045223,PhysRevB.96.115147}; most have been performed within the independent-quasiparticle approximation (IQA), \textit{i.e.}~including self-energy effects to the independent-particle picture.
Beyond this approach, only few studies have reported the impact of MB interactions~\cite{PhysRevB.65.035205,PhysRevB.71.195209,hubener2,PhysRevB.82.235201}.

\begin{figure}
 \includegraphics[width=\linewidth]{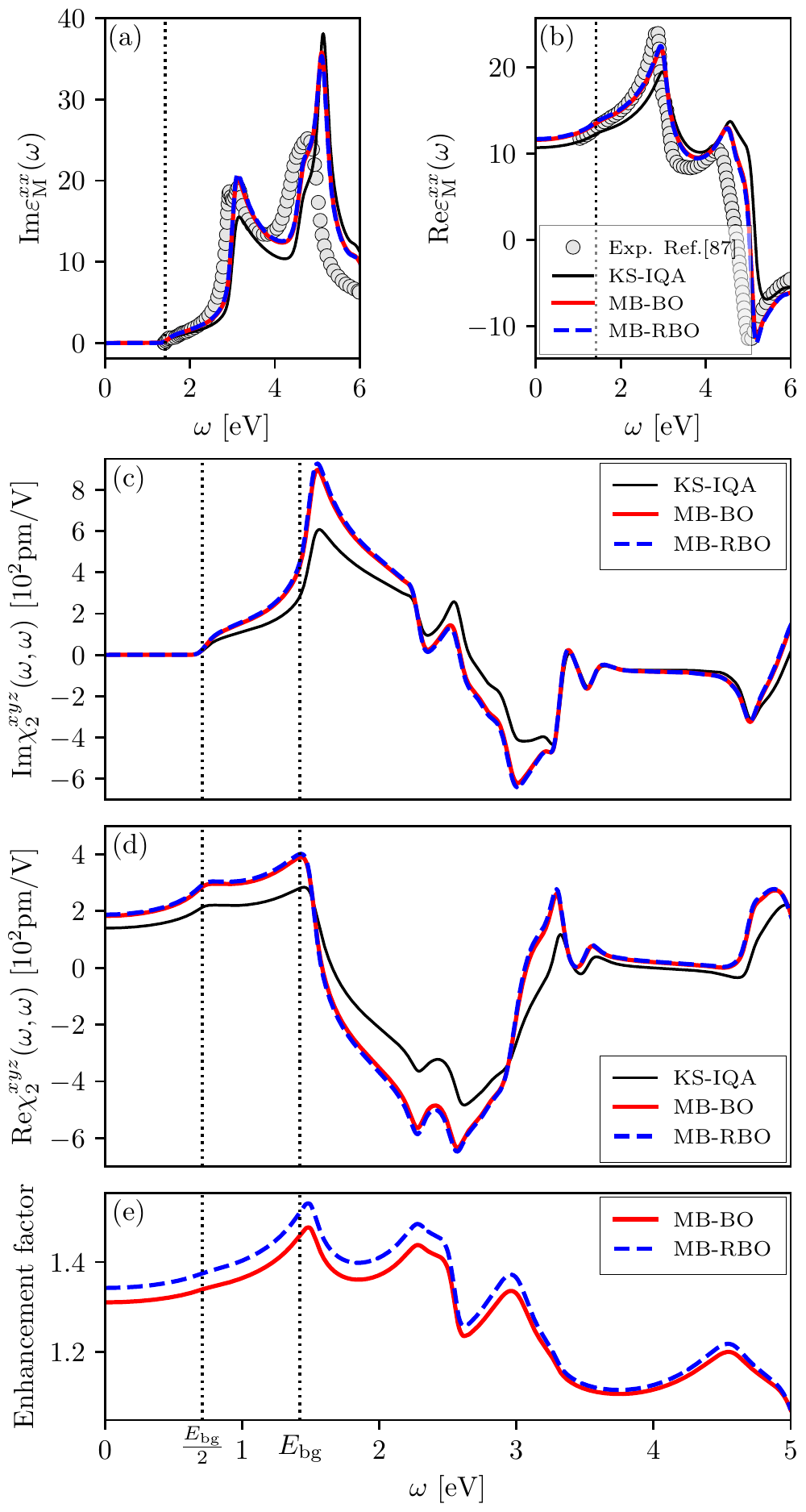}
 \caption{(a) Imaginary and (b) real parts of the macroscopic dielectric function and (c) imaginary and (d) real parts of the macroscopic SHG photosusceptibility for bulk GaAs. Thin solid black, thick solid red and thick dashed blue lines represent the KS-IQA, MB-BO and MB-RBO spectra, respectively. Grey cirlces represent the experimental data from Ref.~\cite{PhysRevLett.90.036801}. Vertical dotted lines represent the band-gap energy $(E_{\mathrm{bg}}=1.42~\mathrm{eV})$ and half its value $(E_{\mathrm{bg}}/2)$. (e) Absolute value of the enhancement factor $\beta^{xyz}(\omega)$ (see Eq.~\ref{eq:beta}).}
 \label{gaas_shg_im_re}
\end{figure}

Since GaAs is a cubic crystal, \mbox{$\varepsilon^{aa}(\omega)=\varepsilon^{xx}(\omega)$} for any Cartesian component $a$, and \mbox{$\chi^{abc}_{2}(\omega,\omega)=\chi^{xyz}_{2}(\omega,\omega)$} for any permutation $abc$ of $xyz$, while all other components of both tensors vanish by symmetry~\cite{boyd2008nonlinear}.
Figures~\hyperref[gaas_shg_im_re]{2(a)} and \hyperref[gaas_shg_im_re]{2(b)} show the spectra of the imaginary and real parts, respectively, of the calculated macroscopic optical dielectric tensor.
Figures~\hyperref[gaas_shg_im_re]{2(c)} and \hyperref[gaas_shg_im_re]{2(d)} show the spectra of the imaginary and real parts, respectively, of the calculated macroscopic SHG photosusceptibility.
KS calculations have been performed within IQA incorporating quasiparticle corrections by means of a scissors operator that rigidly shifts the conduction bands by $0.91~\mathrm{eV}$ in order to recover the experimental value of the band-gap energy at room temperature, \mbox{$E_{\mathrm{bg}}=1.42~\mathrm{eV}$}~\cite{PhysRevB.45.1638}.
In the MB picture, excitonic effects have been included through the LRC xc coefficient \mbox{$\alpha^{a}_{\mathrm{LRC}}\equiv\alpha_{\mathrm{LRC}}$}, which is isotropic in cubic crystals.
The calculated coefficients within BO and RBO approximations are \mbox{$\alpha^{\mathrm{BO}}_{\mathrm{LRC}}=-0.11$} and \mbox{$\alpha^{\mathrm{RBO}}_{\mathrm{LRC}}=-0.12$}, respectively, consistent with the values of previous \textit{ab initio} studies~\cite{PhysRevB.87.205143,PhysRevB.95.205136,doi:10.1063/1.5126501}.
From Figs.~\hyperref[gaas_shg_im_re]{2(a)} and \hyperref[gaas_shg_im_re]{2(b)}, the inclusion of LRC coefficients significantly improves the agreement of linear optics with experimental measurements~\cite{PhysRevLett.90.036801}.

Coming next to quadratic SHG optics, let us  begin by describing the KS-IQA results.
The spectrum of the imaginary part [Fig.~\hyperref[gaas_shg_im_re]{2(c)}] is finite for energies above $E_{\mathrm{bg}}/2$~\cite{PhysRevB.61.5337} and contains a strong peak near the band edge.
As for the real part [Fig.~\hyperref[gaas_shg_im_re]{2(d)}], it is finite at all energies owing to photons absorbed or emitted in virtual excitations.
The spectrum grows progressively at low energies and exhibits maxima at $E_{\mathrm{bg}}/2$ and $E_{\mathrm{bg}}$ due to two- and one-photon resonances, respectively.
At higher energies double resonant transitions take place~\cite{PhysRevB.61.5337} and the spectrum shows several strong peaks.

The net effect of MB-LRC corrections is to increase the magnitude of both the imaginary and real parts of the SHG spectrum, as is clearly visible in Figs.~\hyperref[gaas_shg_im_re]{2(c)} and ~\hyperref[gaas_shg_im_re]{2(d)}, respectively.
The enhancement factor displayed in Fig.~\hyperref[gaas_shg_im_re]{2(e)}
shows that the difference ranges between $0$ and $50~\%$, with the largest renormalization taking place right at the band-edge energy. 
No new spectral feature is formed as a consequence of excitonic effects.
It is also worth noting that the MB-BO and MB-RBO spectra are practically indistinguishable from each other at both first and second orders;
hence, in the remaining of this work we will only show MB-BO results for conciseness.

\begin{figure}\label{gaas_shg_abs}
 \includegraphics[width=\linewidth]{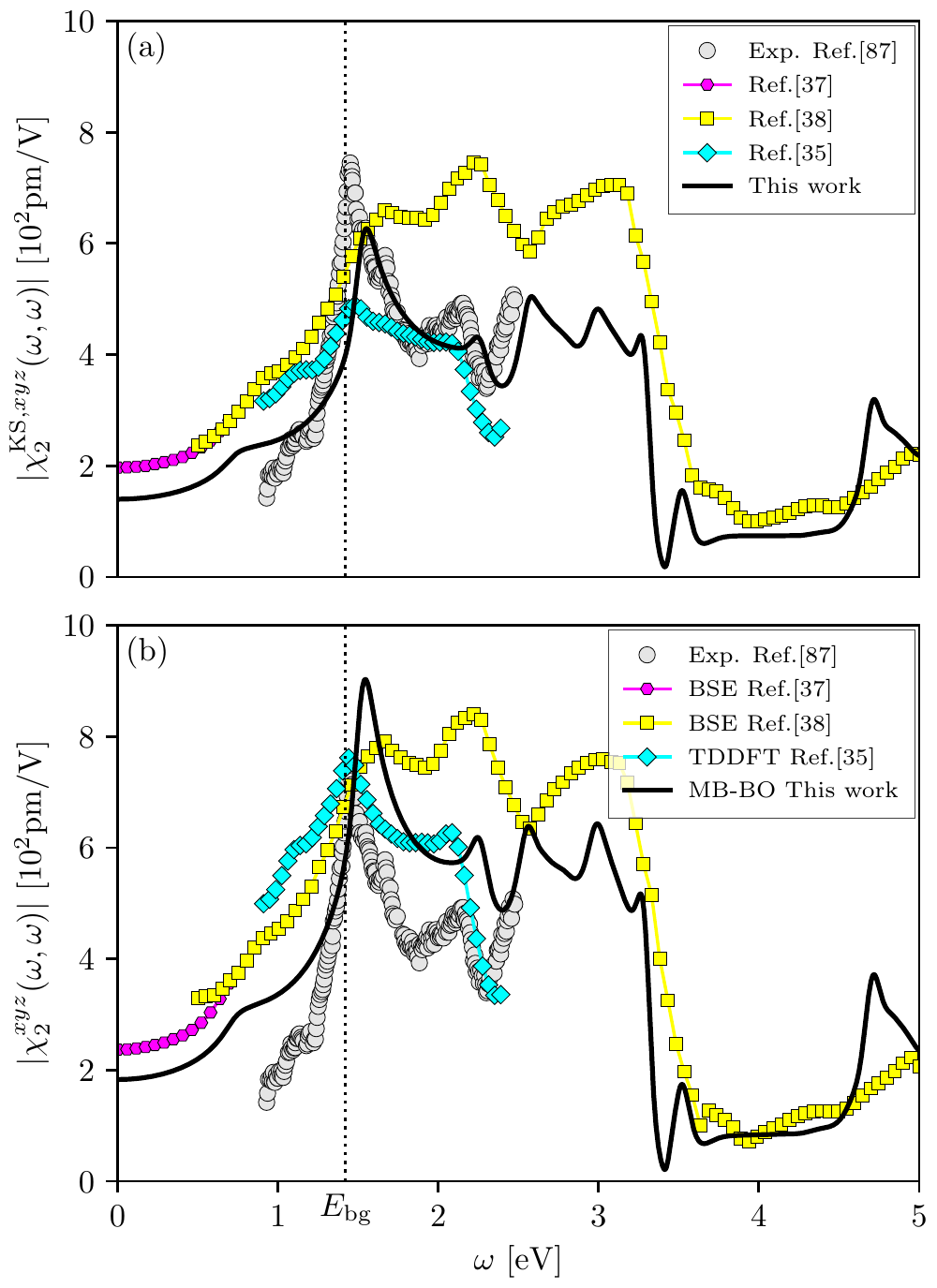}
 \caption{Absolute value of the macroscopic SHG photosusceptibility for bulk GaAs in the (a) KS-IQA and (b) MB-BO pictures. Solid black lines represent our calculated spectra. Grey circles represent the experimental data from Ref.~\cite{PhysRevLett.90.036801}. Magenta hexagons, yellow squares and cyan diamonds represent theoretial spectra from Refs.~\cite{PhysRevB.65.035205}, \cite{PhysRevB.71.195209} and \cite{PhysRevB.82.235201}, respectively, within IQA in (a) and including excitonic effects in (b) by means of BSE for Refs.~\cite{PhysRevB.65.035205} and \cite{PhysRevB.71.195209}, and TDDFT for Ref.~\cite{PhysRevB.82.235201}. In all these works, the scissors shift is such that the experimental value of the band-gap energy is recovered, being equal to $0.6~\mathrm{eV}$ and $0.8~\mathrm{eV}$ for Refs.~\cite{PhysRevB.71.195209} and \cite{PhysRevB.82.235201}, respectively. In Ref.~\cite{PhysRevB.82.235201}, TDDFT is employed with an empirical $\alpha_{\mathrm{LRC}}=0.2$.}
 \label{gaas_shg_abs}
\end{figure}
In Fig.~\ref{gaas_shg_abs} we show the absolute value of the macroscopic SHG photosusceptibility and compare our calculations with experimental measurements as well as previous theoretical works including different approximations.
The experimental spectrum is dominated by a peak at the band-edge energy and contains a ``V''-shaped form  between $2$ and $3~\mathrm{eV}$.
These two spectral features are well described by both our KS-IQA and MB-LRC calculations, which show similar shape but different size as discussed previously.
Our KS-IQA result [see Fig.~\hyperref[gaas_shg_abs]{3(a)}] is in qualitative agreement with previous IQA calculations, specially that of Ref.~\cite{PhysRevB.82.235201}.
Our MB-LRC calculation [see Fig.~\hyperref[gaas_shg_abs]{3(b)}] strikes the best balance in describing the magnitude and width of the two spectral features of the experiment, although the height of the ``V''-shaped form is somewhat overestimated.
Here too we note a qualitative agreement with the TDDFT result of Ref.~\cite{PhysRevB.82.235201}. 

In quantitative terms, our results show sharper peaks than those of previous theoretical works.
This can be a consequence of the small smearing factors achieveable thanks to Wannier interpolation, which makes it possible to consider on the order of $10^{6}$ $k$-points for converging the SHG integrals over the BZ (see Eqs.~\ref{alphatershg} and \ref{sigmatrashg}).
For comparison, the calculations of Refs.~\cite{PhysRevB.71.195209} and~\cite{PhysRevB.82.235201} employed on the order of $10^{3}$ and $10^{4}$ $k$-points, respectively.
This fine sampling has allowed us to model the lifetime of hot carriers ($\sim10~\mathrm{fs}$) in bulk GaAs~\cite{doi:10.1073/pnas.1419446112}, which therefore renders more realistic spectral widths as compared to experiment.
\subsection{BC$_{2}$N}
The graphitic-layered semiconductor BC$_{2}$N has attracted interest in the last years as a potential nonlinear optical material~\cite{PhysRevLett.77.187,PhysRevLett.83.2406}.
Its layered geometry composed of alternating zigzag of C$-$C and B$-$N chains makes it a malleable and strongly anisotropic crystal~\cite{PhysRevB.39.1760}.
Among its several polytypes, the A2 configuration (BC$_{2}$N-A2) is the most stable noncentrosymmetric structure~\cite{PhysRevB.73.193304} that allows a finite quadratic response.
First-principles calculations within the independent-particle approximation (IPA) have recently predicted a large SHG for BC$_{2}$N-A2 in monolayer and nanotube form~\cite{Lucking2018} that is an order of magnitude larger than in bulk GaAs.
A large shift current has also been calculated recently in bulk~\cite{PhysRevResearch.2.013263} and monolayer~\cite{10.21468/SciPostPhys.12.2.070} form.
To our knowledge, no systematic study of MB effects on the SHG has been carried out for bulk BC$_{2}$N-A2 up to date.
\begin{figure}
 \includegraphics[width=\linewidth]{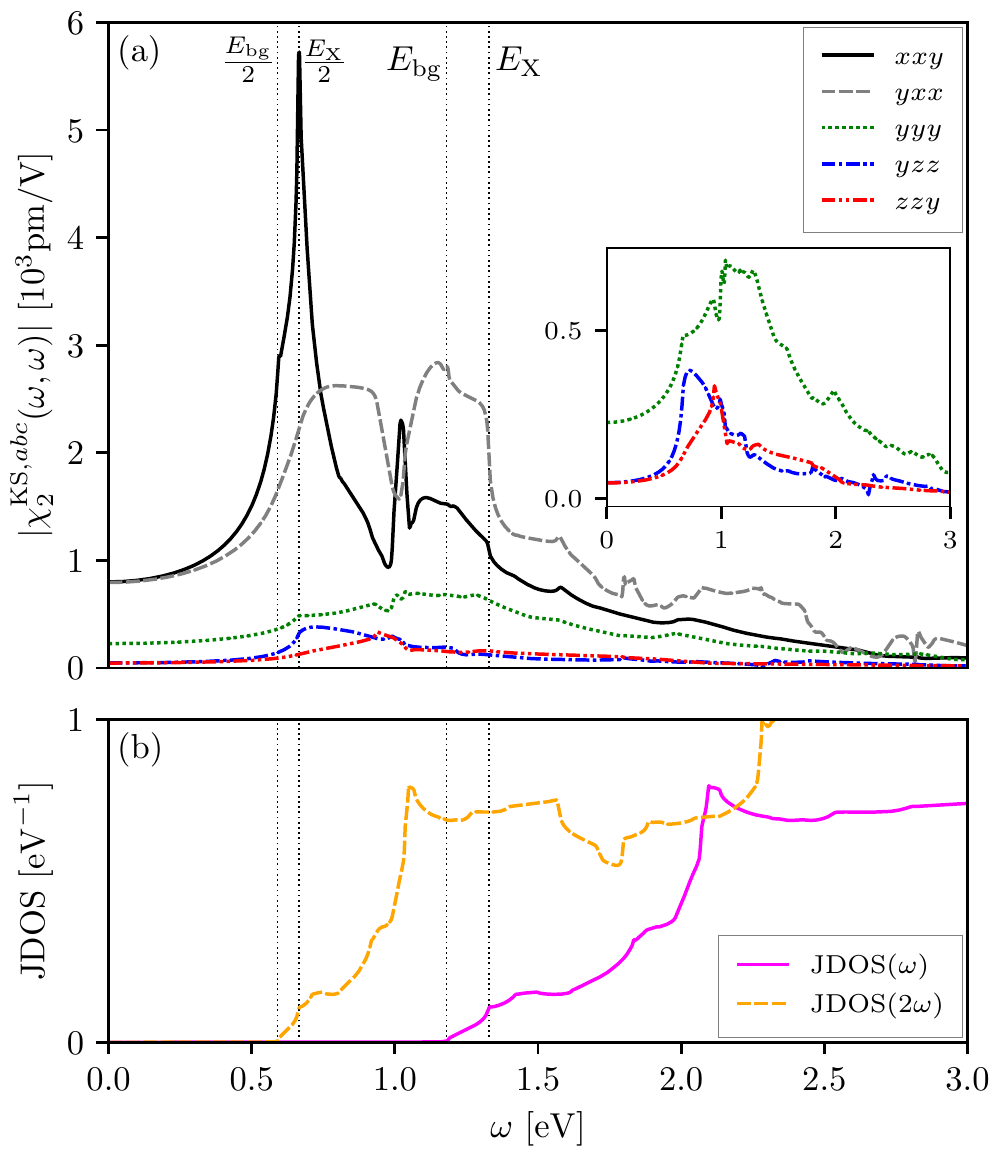}
 \caption{(a) Absolute value of the SHG KS-IPA photosusceptibility tensor for bulk BC$_{2}$N-A2. Solid black, dashed grey, dotted green, dashdotdashed blue and dashdotdotted red lines represent the spectra of the $xxy=xyx$, $yxx$, $yyy$, $yzz$ and $zzy=zyz$ non-vanishing components, respectively. The inset zooms in the $yyy$, $yzz$ and $zzy$ components. (b) Joint density of states. Solid magenta and dashed orange lines represent the one- and two-photon signals, respectively. Vertical dotted lines represent the band-edge energy range boundaries $(E_{\mathrm{bg}}\approx1.18$ and $E_{\mathrm{X}}\approx1.33)$ and half their values \sout{$(E_{\mathrm{bg}}/2$ and $E_{\mathrm{X}}/2)$}.}
 \label{bc2n_shg0}
\end{figure}

Owing to its \textit{mm2} point group, the symmetry-allowed components of the SHG photosusceptibility tensor for BC$_{2}$N-A2 are $xxy=xyx$, $yxx$, $yyy$ $yzz$ and $zzy=zyz$~\cite{boyd2008nonlinear}.
Their absolute values in the KS-IPA picture are displayed in Fig.~\hyperref[bc2n_shg0]{4(a)}.
In order to facilitate the discussion of the spectral features, in Fig.~\hyperref[bc2n_shg0]{4(b)} we show the joint density of states (JDOS) per crystal unit cell~\cite{PhysRevResearch.2.013263} for the one- and two-photon signals.
In these and following figures, $E_{\mathrm{bg}}=1.18~\mathrm{eV}$ denotes the direct band-gap energy, while $E_{\mathrm{X}}=1.33~\mathrm{eV}$ refers to the band-gap energy at high symmetry point $\mathrm{X}$.
The latter was found to mark the peak absorption of the shift current at low energies~\cite{PhysRevResearch.2.013263} and will also play an important role in the SHG.

The tensor components $xxy$ and $yxx$ dominate the SHG photoresponse with values of the order of the SHG for the monolayer and nanotube forms~\cite{Lucking2018}, and coincide with the dominant components of the shift current for bulk BC$_{2}$N-A2~\cite{PhysRevResearch.2.013263}.
The maximum value of \mbox{$5.8\times10^{3}~\mathrm{pm/V}$} takes place for the $yxx$ component at $E_{\mathrm{X}}/2$ owing to a two-photon absorption process.
These and further specral features like the peak at $\simeq1~\mathrm{eV}$ can be associated to contributions in the one- and two-photon JDOS [see Fig.~\hyperref[bc2n_shg0]{4(b)}].
\begin{figure}
 \includegraphics[width=\linewidth]{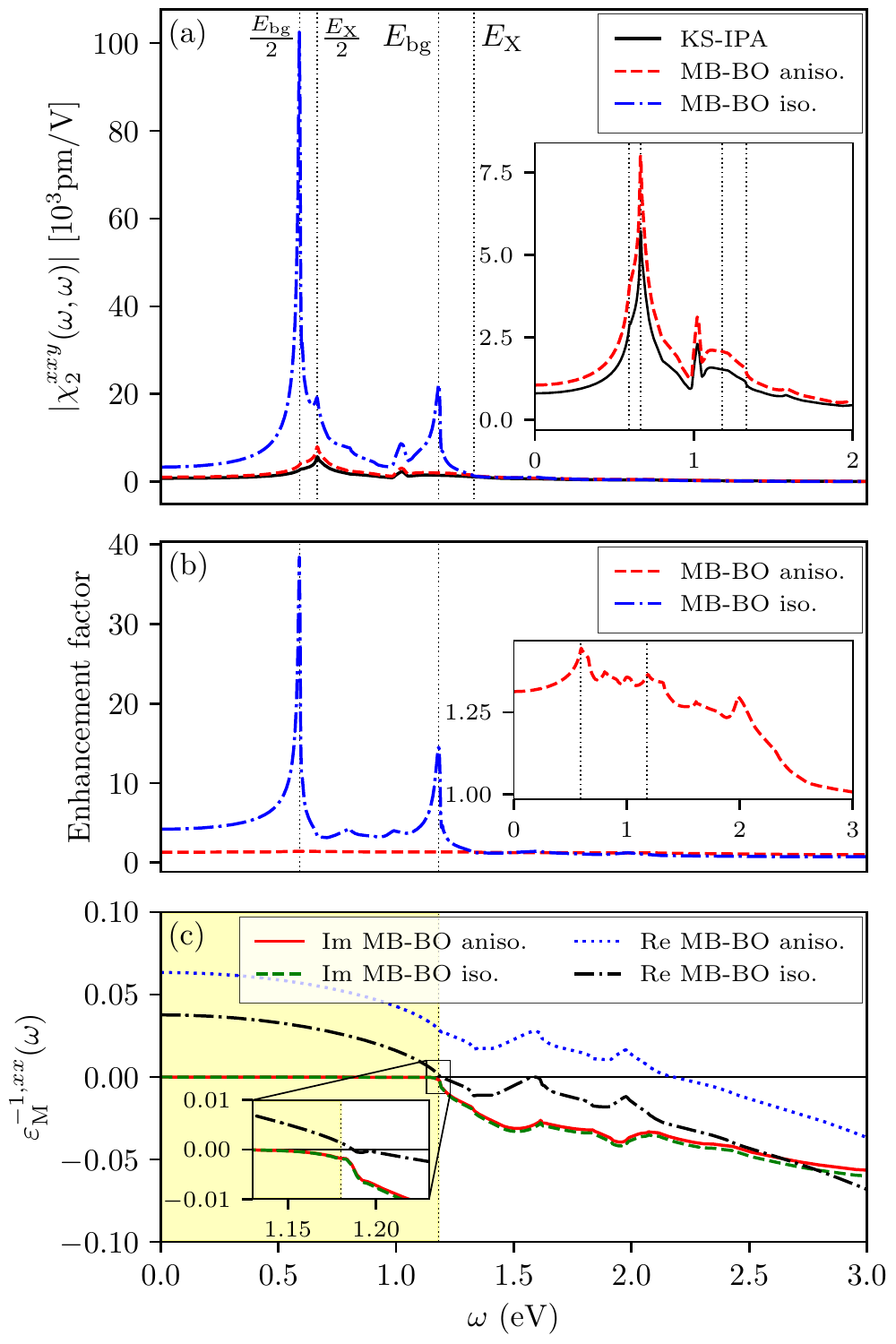}
 \caption{(a) Absolute value of the $xxy$ component of the macroscopic SHG photosusceptibility tensor for bulk BC$_{2}$N-A2. (b) Absolute value of the enhancement factor $\beta^{xxy}(\omega)$. The solid black line represent the KS-IPA spectrum. The dashed red and dashdotted blue lines represent the MB-BO spectrum using the anisotropic (aniso.) and isotropic (iso.) tensorial LRC xc kernel, respectively. (c) $xx$ component of the inverse of the macroscopic optical dielectric tensor. The solid red (dashed green) and dotted blue (dashdotted black) lines represent the real (Re) and imaginary (Im) parts in the anisotropic (isotropic) case, respectively. These spectra are practically identical when using BO or RBO approximations.}
 \label{bc2n_shg_xxz}
\end{figure}
\begin{figure*}[t]
 \includegraphics[width=\linewidth]{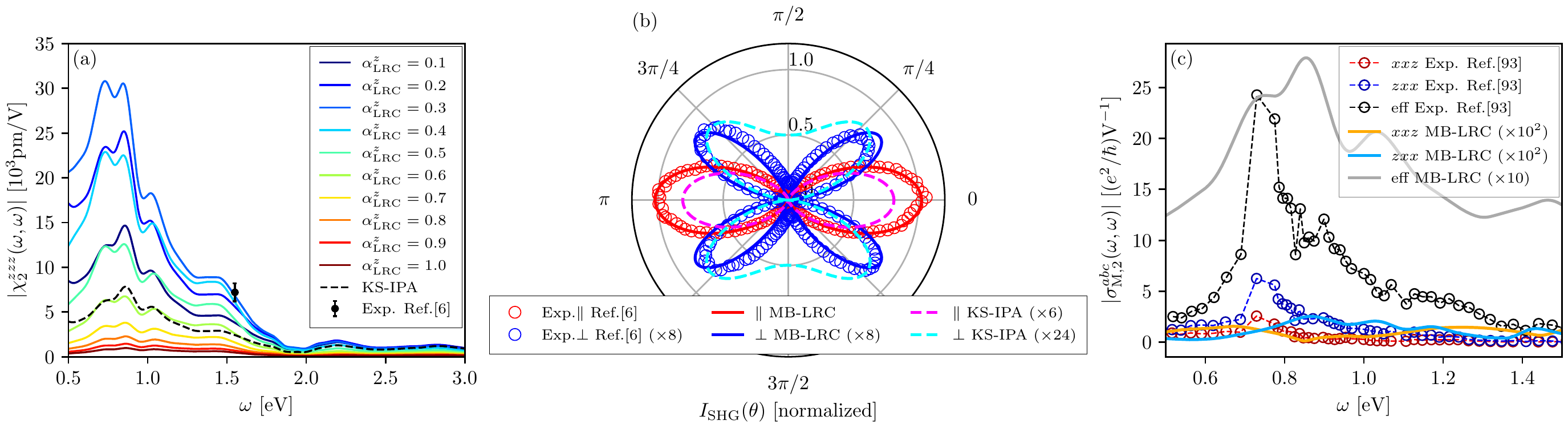}
 \caption{(a) Absolute value of the $zzz$ component of the macroscopic SHG photosusceptibility tensor for bulk TaAs. Dashed black and solid colored lines represent the KS-IPA and MB-LRC spectra as a function of $\alpha^{z}_{\mathrm{LRC}}$, respectively. Black errorbar corresponds to the experimental datapoint from Ref.~\cite{Wu2017} (b) SHG intensity polar plot in both parallel $(\parallel)$ and perpendicular $(\perp)$ generator/analyser configurations. For better visualization, results in the $\perp$ configuration are multiplied by a factor $8$. KS-IPA calculations are multiplied by a factor $6$ and $24$ for $\parallel$ and $\perp$ configurations, respectively. Open red and blue circles represent $\parallel$ and $\perp$ experimental data from Ref.~\cite{Wu2017}, respectively. Solid (dashed) red (magenta) and blue (cyan) lines represent our MB-LRC (KS-IPA) calculations in the $\parallel$ and $\perp$ configurations, respectively, for \mbox{$\{\alpha^{x=y}_{\mathrm{LRC}}=1.8,\alpha^{z}_{\mathrm{LRC}}=0.3\}$}. (c) Absolute value of the macroscopic SHG photoconductivity tensor. Open red, dark blue and black circles represent experimental data from Ref.~\cite{PhysRevB.98.165113} for the $xxz$, $zxx$ and effective components, respectively. Orange, cyan and grey solid lines represent our calculated MB-LRC spectra of the $xxz$, $zxx$ (multiplied by 100) and effective (multiplied by 10) components, respectively.}
 \label{fig:taas}
\end{figure*}

We focus next on the MB-LRC interactions.
Unlike the case of GaAs studied previously, BC$_{2}$N-A2 is anisotropic and so is the tensorial LRC xc kernel;
the calculated  BO coefficients are \mbox{$\{\alpha_{\mathrm{LRC}}^{x},\alpha_{\mathrm{LRC}}^{y},\alpha_{\mathrm{LRC}}^{z}\}=\{0.06,0.29,5.51\}$}.
Note that the $z$ component is an order of magnitude larger than the $x$ and $y$ components, as well as the coefficient computed for GaAs (see Sec.~\ref{subsec:gaas}).
Therefore, BC$_{2}$N-A2 represents a clear example where an isotropic treatment of the excitonic effects constitutes a poor choice, given that the value of the space-averaged scalar LRC xc coefficient \mbox{$\alpha^{\mathrm{iso.}}_{\mathrm{LRC}}=0.42$} is close to none of the actual space-resolved tensorial components.
In the following, we illustrate the profound errors that this procedure can induce in the absorption spectrum.

In Fig.~\hyperref[bc2n_shg_xxz]{5(a)} we display the renormalization of the macroscopic SHG photosusceptibility tensor component $xxy$ by electron-hole corrections at two levels: using the anisotropic and isotropic tensorial LRC xc kernels.
Comparison to the KS-IPA response shows that the anisotropic kernel induces a maximum increase of nearly a factor $1.5$ [see enhancement factor in Fig.~\hyperref[bc2n_shg_xxz]{5(b)}], but does not alter the overall shape of the spectrum, in line with what we found for GaAs (see Sec.~\ref{subsec:gaas}).
On the other hand, the isotropic kernel produces a large peak at half the band-edge energy that completely dominates the MB-BO spectrum, with an enhancement of more than one order of magnitude as compared to the anisotropic kernel.
A secondary peak is also visible at the band-edge energy.

The origin of these two sharp peaks can be determined by inspecting the
inverse of the macroscopic optical dielectric tensor along $x$; this quantity is shown in Fig.~\hyperref[bc2n_shg_xxz]{5(c)} separately for the real and imaginary parts.
While \mbox{$\mathrm{Im}\epsilon_{\mathrm{M}}^{-1,xx}(\omega)$} is barely affected by the type of tensorial LRC xc kernel, \mbox{$\mathrm{Re}\epsilon_{\mathrm{M}}^{-1,xx}(\omega)$} shows a strong shift that is nearly frequency-independent; both these features can be qualitatively understood by working out explicit expressions (use Eqs.~\ref{eq:epsmbs} and~\ref{eq:sigma1} in Eq.~\ref{eq:epsM}) and noting that the Hartree contribution is much stronger than any of the LRC xc components, \textit{i.e.}, \mbox{$4\pi\gg\alpha^{a}_{\mathrm{LRC}}$}.
In the case of the isotropic kernel, \mbox{$\mathrm{Re}\epsilon_{\mathrm{M}}^{-1,xx}(\omega)$} crosses the zero axis very close to the band-edge energy, where $\mathrm{Im}\epsilon_{\mathrm{M}}^{-1,xx}(\omega)\simeq0$ too, leading to a sharp peak in $\epsilon^{xx}_{\mathrm{M}}(\omega)$ at that energy [see enhancement factor in Fig.~\hyperref[bc2n_shg_xxz]{5(b)}].
This peak is then replicated at half the band-edge energy in the SHG spectrum through the $\epsilon^{xx}_{\mathrm{M}}(2\omega)$ factor in Eq.~\ref{micro2macroshg}, and enhanced by transition matrix-elements.

We have verified that a similar effect takes place for the SHG tensor component $yxx$ too (not shown).
In this case, the isotropic kernel gives rise to a even larger peak right at the band-edge energy reaching $\simeq700\times10^{3}~\mathrm{pm/V}$ (see Fig.~\hyperref[bc2n_shg_xxz]{5(a)} for comparison), while the anisotropic kernel induces only moderate changes to the KS-IPA response.
These examples show that sharp, exciton-like peaks in the SHG spectrum can be induced by MB-LRC effects provided the appropriate conditions are met.
These conditions are very sensitive to numerics, which stresses the importance of accounting for the space-resolved anisotropy of the material in the tensorial xc kernel, and therefore, its advantage over a space-averaged scalar approach.
\subsection{TaAs}\label{subsec:taas}
Theoretically predicted~\cite{Huang2015,PhysRevX.5.011029} 
and experimentally confirmed in 2015~\cite{PhysRevX.5.031013,doi:10.1126/science.aaa9297,Lv2015}, TaAs is a type I Weyl semimetal~\cite{Weyl1929} without an inversion center.
Following its discovery, several experiments have reported remarkable nonlinear optical properties.
Ref.~\cite{Wu2017} measured a ``giant'' SHG photosusceptibility at $\simeq1.55$ eV that is an order of magnitude larger than in most other materials.
Shortly after, Ref.~\cite{PhysRevB.98.165113} extended the measurements to lower energies and found a narrow resonance at $\simeq0.75~\mathrm{eV}$ with an even larger photoresponse.
In addition to the SHG, other quadratic optical responses such as the
shift current have also been measured to be exceptionally large~\cite{Osterhoudt2019}.

Due to its $4mm$ point group, the symmetry-allowed components of the SHG tensor in TaAs are $zzz$, $zxx=zxz=zyy=zyz$ and $xxz=xzx=yyz=yzy$, where $x$ and $y$ are equivalent directions of the tetragonal unit cell and the direction perpendicular to the $xy$ plane is the polar axis $z$.

Unlike GaAs and BC$_{2}$N studied previously, TaAs is a semimetal.
In this case,  the BO and RBO approximations for the calculation of the tensorial LRC xc kernel cannot be applied directly since $\mathrm{Im}[\epsilon(\omega=0)]\neq0$~\cite{PhysRevLett.107.186401}.
In consequence, we have chosen to renormalize the SHG KS-IPA spectrum for a reasonable range of LRC xc coefficients \mbox{$\{\alpha_{\mathrm{LRC}}^{x=y},\alpha^{z}_{\mathrm{LRC}}\}$} and determine empirically their most appropriate values by comparing to the experimental measurements.

In Fig.~\hyperref[fig:taas]{6(a)} we show our calculated $|\chi^{zzz}_{2}(\omega,\omega)|$ as a function of \mbox{$\alpha_{\mathrm{LRC}}^{z}$} together with the available experimental datapoint at \mbox{$\simeq 1.55~\mathrm{eV}$} from Ref.~\cite{Wu2017}, equal to $7\pm1\times10^{3}~\mathrm{pm/V}$.
The KS-IPA response peaks around $0.85~\mathrm{eV}$, and captures the magnitude of the experimental value but underestimates it by roughly a factor two.
The SHG MB-LRC spectrum grows with the value of \mbox{$\alpha_{\mathrm{LRC}}^{z}$} until it equals $0.3$, where it basically matches the experiment and therefore represents the optimal value.
For \mbox{$\alpha_{\mathrm{LRC}}^{z}>0.3$}, the magnitude of \mbox{$|\chi^{zzz}_{2}(\omega,\omega)|$} starts decreasing and it becomes nearly overdamped for \mbox{$\alpha_{\mathrm{LRC}}^{z}>0.6$}.
The overall shape of the spectrum is maintained in the whole range of $\alpha_{\mathrm{LRC}}^{z}$ considered.
By applying the same procedure to the $zxx$ and $xxz$ components we have determined the remaining coefficient \mbox{$\alpha_{\mathrm{LRC}}^{x=y}=1.8$}.

In Ref.~\cite{Wu2017}, two additional measurements were conducted at \mbox{$\simeq1.55~\mathrm{eV}$} for varying angle $\theta$ of linearly-polarized light, with the field oriented along the \mbox{[1,1,-1]} (parallel setup, $\parallel$) and \mbox{[1,-1,0]} (perpendicular setup, $\perp$) directions.
Making use of the appropriate combination of the SHG tensor components (see Eqs.~3 and 4 of the Supplementary Information in Ref.~\cite{Wu2017}), we have calculated the angular dependence of the SHG intensity and compared it to the experimental polar plot, as shown in Fig.~\hyperref[fig:taas]{6(b)}.
For the parallel configuration, the response shows an elongated shape along the $\theta=0$ axis that is remarkably well captured by our MB-LRC result.
For the perpendicular configuration, the response shows a four-fold structure with maxima at $\pi/4+n\cdot\pi/2$ and minima at $n\cdot\pi/2$ for any integer $n$.
While the KS-IPA calculation fails in both magnitude and shape, our MB-LRC result nicely agrees with the experimental measurement, thus capturing the main characteristics of the photoresponse at this particular energy.

As the last step, we proceed to study the low-energy region accessed in Ref.~\cite{PhysRevB.98.165113}, where a narrow resonance was measured at $\simeq0.75$ eV.
In Fig.~\hyperref[fig:taas]{6(c)} we compare the experimentally measured \mbox{$|\sigma^{zxx}_{\mathrm{M},2}|$}, \mbox{$|\sigma^{xxz}_{\mathrm{M},2}|$} and \mbox{$|\sigma^{\mathrm{eff}}_{\mathrm{M},2}|\equiv|\sigma^{zzz}_{\mathrm{M},2}+4\sigma^{xxz}_{\mathrm{M},2}+2\sigma^{zxx}_{\mathrm{M},2}|$} with our calculations using the optimal values of $\alpha_{\mathrm{LRC}}^{a}$ quoted previously.
Our results underestimate the main exciton-like peak by an order of magnitude, and we have been unable to strike a substantial improvement by further varying $\alpha_{\mathrm{LRC}}^{a}$.
The description of this low-energy peak appears therefore 
to be beyond the scope of the linear tensorial LRC xc kernel considered here.
It is tempting to speculate that it might be induced by MB corrections not included in our calculations, \textit{e.g.}, a frequency dependence in the LRC xc coefficients $\alpha^{a}_{\mathrm{LRC}}(\omega)$~\cite{PhysRevB.67.045207,PhysRevB.72.125203}, or the quadratic tensorial xc kernel of Eq.~\ref{exc}.
\section{SUMMARY AND OUTLOOK}\label{sec:conclusions}
In summary, we have described a general scheme for calculating the quadratic optical response to light tensor of crystals taking into account many-body interactions.
We have formally included excitonic effects by means of a tensorial long-range exchange-correlation kernel whose coefficients have been calculated using 
two variants of the parameter-free bootstrap approximation.
We have also generalized previous
independent-particle expressions~\cite{PhysRevB.48.11705,PhysRevB.52.14636,PhysRevB.61.5337,PhysRevResearch.2.012017,PhysRevResearch.3.L042032} for the transition matrix elements to account for all metallic contributions, allowing an exhaustive study of materials like Weyl semimetals.

Linking the formalism with the Wannier interpolation of the transition matrix elements~\cite{RevModPhys.84.1419,PhysRevB.74.195118,PhysRevB.97.245143}, we have performed calculations of the second-harmonic generation photoresponse tensor in a range of materials.
Besides benchmarking our approafch in bulk GaAs, we have shown that the electron-hole attraction can give rise to strong and sharply localized one- and two-photon resonances that are absent in the Kohn-Sham photoresponse.
In the graphitic-layered bulk crystal BC$_2$N, an space-averaged isotropic approach overestimates the electronic renormalization by orders of magnitude, highlighting the need of accounting for the space-resolved anisotropic nature of many-body interactions in tensorial form.
We have further verified that the bootstrap and the RPA-bootstrap kernels yield virtually the same result, consistent with previous studies in small to medium-gap semiconductors~\cite{PhysRevLett.117.159701,PhysRevLett.117.159702,doi:10.1063/1.5126501}.
Finally, with the use of a highly dense $k$-space mesh, our calculations have  reproduced the magnitude and angular dependence of the photoresponse for the Weyl semimetal TaAs measured recently~\cite{Wu2017}.

We hope that the presented scheme together with its implementation in the {\tt Wannier90} and {\tt WannierBerri} code packages will facilitate an  efficient and accurate calculation of the quadratic optical photoresponse of materials beyond the SHG process analyzed here.
We note that the procedure adopted for including many-body excitonic effects requires only a fraction of the computational time as compared to the calculation of the Kohn-Sham photoresponse.

The proposed method can be improved in several fronts.
Adopting a Wannier-based strategy for the calculation of the linear xc kernel in metals and semimetals (see \textit{e.g.}, Ref.~\cite{PhysRevB.85.054305}) would allow a 
fully parameter-free analysis in these type of materials.
An improved description of many-body effects can be achieved by extending the LRC xc coefficients to frequency domain~\cite{PhysRevB.67.045207,PhysRevB.72.125203} or by working out an approximation for the second-order xc kernel, which would open the way to study potentially new excitonic effects that have been barely described in the literature up to now.
The method can also model crystal local-field corrections, whose effect
tends to reduce the intensity of the SHG spectra~\cite{PhysRevLett.66.41}
and could therefore improve agreement with experiments.
Finally, accounting for quasiparticle self-energy corrections due to electron-electron or electron-phonon interactions would allow modelling extrinsic quadratic contributions such as the ballistic current~\cite{sturman-book92,PhysRevB.104.235203,PhysRevLett.126.177403,Sturman_2020}.
We expect to address these subjects in future works.
\section{ACKNOWLEDGMENTS}
We are very grateful to Ivo Souza and Fernando de Juan for helpful discussions.
This project has received funding from the European Union's Horizon 2020 research and innovation programme under the European Research Council (ERC) grant agreement No 946629.  
\hspace{1cm}\\
\appendix
\section{DERIVATION OF THE QUADRATIC DYSON-LIKE RESPONSE TENSOR EQUATION}\label{appxjjj}
Here we outline the steps involved in the derivation of the Dyson-like equation relating the MB and KS conductivity tensors at second order in Eq.~\ref{sigma2de} of the main text.
We start by applying the chain rule twice in the definition of the quadratic MB conductivity tensor in Eq.~\ref{sigma2},
\begin{widetext}
 \begin{equation}\label{appsigma1}
  \begin{split}
   \sigma_{1}^{abc}(1,2,3)=&\frac{\delta}{\delta{E_\mathrm{ext}^{b}}(2)}\left[\int\sum_{d}\frac{\delta J^{a}(1)}{\delta{E_\mathrm{tot}^{d}}(4)}\frac{\delta{E_\mathrm{tot}^{d}}(4)}{\delta{E_\mathrm{ext}^{c}}(3)}d4\right]\\
   =&\int\sum_{d}\left\{\frac{\delta^{2}J^{a}(1)}{\delta{E_\mathrm{ext}^{b}}(2)\delta{E_\mathrm{tot}^{d}}(4)}\frac{\delta{E_\mathrm{tot}^{d}}(4)}{\delta{E_\mathrm{ext}^{c}}(3)}+\frac{\delta J^{a}(1)}{\delta{E_\mathrm{tot}^{d}}(4)}\frac{\delta}{\delta{E_\mathrm{ext}^{b}}(2)}\left[\frac{\delta{E_\mathrm{tot}^{d}}(4)}{\delta{E_\mathrm{ext}^{c}}(3)}\right]\right\}d4\\
   =&\iint\sum_{de}\frac{\delta^{2}J^{a}(1)}{\delta{E_\mathrm{tot}^{e}}(5)\delta{E_\mathrm{tot}^{d}}(4)}\frac{\delta{E_\mathrm{tot}^{e}}(5)}{\delta{E_\mathrm{ext}^{b}}(2)}\frac{\delta{E_\mathrm{tot}^{d}}(4)}{\delta{E_\mathrm{ext}^{c}}(3)}d4d5+\int\sum_{d}\frac{\delta J^{a}(1)}{\delta{E_\mathrm{tot}^{d}}(4)}\frac{\delta^{2}{E_\mathrm{tot}^{d}}(4)}{\delta{E_\mathrm{ext}^{b}}(2)\delta{E_\mathrm{ext}^{c}}(3)}d4.
  \end{split}
 \end{equation}
The first term on the right-hand side (r.h.s.)~of the last line in Eq.~\ref{appsigma1} can be expressed in terms of $\overline{\sigma}^{\mathrm{KS}}_{2}$ and $\epsilon$ using Eqs.~\ref{sigma20} and~\ref{etot2eext}, respectively.
As for the second term, the piece ${\delta J^{a}(1)}/{\delta{E_\mathrm{tot}^{d}}(4)}$ can be written in terms of $\overline{\sigma}^{\mathrm{KS}}_{1}$ using Eq.~\ref{sigma10}, while the calculation of the remaining piece requires applying the chain rule again,
 \begin{equation}\label{d2etotdeper2}
  \begin{split}
   \frac{\delta^{2}{E_\mathrm{tot}^{d}}(4)}{\delta{E_\mathrm{ext}^{b}}(2)\delta{E_\mathrm{ext}^{c}}(3)}&=\frac{\delta}{\delta{E_\mathrm{ext}^{b}}(2)}\left[\delta(4,3)\delta_{dc}+\int\sum_{e}\frac{\delta{E_\mathrm{Hxc}^{d}}(4)}{\delta J^{e}(5)}\frac{\delta J^{e}(5)}{\delta{E_\mathrm{ext}^{c}}(3)}d5\right]\\&=\int\sum_{e}\left[\frac{\delta^{2}{E_\mathrm{Hxc}^{d}}(4)}{\delta{E_\mathrm{ext}^{b}}(2)\delta J^{e}(5)}\frac{\delta J^{e}(5)}{\delta{E_\mathrm{ext}^{c}}(3)}+\frac{\delta{E_\mathrm{Hxc}^{d}}(4)}{\delta J^{e}(5)}\frac{\delta^{2} J^{e}(5)}{\delta{E_\mathrm{ext}^{b}}(2)\delta{E_\mathrm{ext}^{c}}(3)}\right]d5\\&=\iint\sum_{ef}\frac{\delta^{2}{E_\mathrm{Hxc}^{d}}(4)}{\delta J^{f}(6)\delta J^{e}(5)}\frac{\delta J^{f}(6)}{\delta{E_\mathrm{ext}^{b}}(2)}\frac{\delta J^{e}(5)}{\delta{E_\mathrm{ext}^{c}}(3)}d5d6+\int\sum_{e}\frac{\delta{E_\mathrm{Hxc}^{d}}(4)}{\delta J^{e}(5)}\frac{\delta^{2} J^{e}(5)}{\delta{E_\mathrm{ext}^{b}}(2)\delta{E_\mathrm{ext}^{c}}(3)}d5,
  \end{split}
 \end{equation}
where we used $\varepsilon^{-1,ab}(1,2)=\frac{\delta E^{a}_{\mathrm{tot}}(1)}{\delta E^{b}_{\mathrm{ext}}(2)}$ and Eq.~\ref{sigma20}.
The first term on the r.h.s.~of the last line in Eq.~\ref{d2etotdeper2} can be expressed in terms of $K^{abc}_{\mathrm{xc},2}(1,2,3)$ and $\sigma^{ab}_{1}(1,2)$ using $\frac{\delta^{2}E_\mathrm{Hxc}^{a}(1)}{\delta J^{b}(2)\delta J^{c}(3)}=K_{\mathrm{xc},2}^{abc}(1,2,3)$ and Eq.~\ref{sigma1}, respectively.
As for the second piece, it can be recast in terms of $K^{ab}_{\mathrm{Hxc},1}(1,2)$ and $\sigma_{2}$ using $K^{ab}_{\mathrm{Hxc},1}(1,2)=\frac{\delta E^{a}_{\mathrm{Hxc}}(1)}{\delta J^{b}(2)}$ and Eqs.~\ref{sigma2}, respectively.

Taking into account all the previous observations, we can rewrite Eq.~\ref{appsigma1} as
 \begin{equation}\label{appsigma2}
  \begin{split}
   \sigma_{2}^{abc}(1,2,3)=&\iint\sum_{de}\sigma_{2}^{\mathrm{KS},aed}(1,5,4)\varepsilon^{-1,eb}(5,2)\varepsilon^{-1,dc}(4,3)d4d5\\&+\iiint\sum_{def}\sigma_{1}^{\mathrm{KS},ad}(1,4)K_{\mathrm{xc},2}^{dfe}(4,6,5)\sigma_{1}^{fb}(6,2)\sigma_{1}^{ec}(5,3)d4d5d6\\&+\iint\sum_{de}\sigma_{1}^{\mathrm{KS},ad}(1,4)K_{\mathrm{Hxc},1}^{de}(4,5)\sigma_{2}^{ebc}(5,2,3)d4d5.
  \end{split}
 \end{equation}
Moving now the last term on the r.h.s.~of Eq.~\ref{appsigma2} to the left-hand side (l.h.s.), we can rewrite this side with the quadratic MB conductivity tensor as a common factor.
Taking advantage from the definition of the dielectric tensor in Eq.~\ref{deextdetot}, we obtain that
 \begin{equation}
  \begin{split}
   &\int\sum_{e}\left[\delta(1,5)\delta_{ae}-\int\sum_{d}\sigma_{1}^{\mathrm{KS},ad}(1,4)K_{\mathrm{Hxc},1}^{de}(4,5)d4\right]\sigma_{2}^{ebc}(5,2,3)d5\equiv\int\sum_{d}\varepsilon^{da}(1,4)\sigma_{2}^{dbc}(4,2,3)d4=\\&\iint\sum_{de}\sigma_{2}^{\mathrm{KS},aed}(1,5,4)\varepsilon^{-1,eb}(5,2)\varepsilon^{-1,dc}(4,3)d4d5+\iiint\sum_{def}\sigma_{1}^{\mathrm{KS},ad}(1,4)K_{\mathrm{xc},2}^{dfe}(4,6,5)\sigma_{1}^{fb}(6,2)\sigma_{1}^{ec}(5,3)d4d5d6.
  \end{split}
 \end{equation}
Finally, inverting the transpose of the dielectric tensor from the r.h.s.~to the l.h.s., we arrive at the Dyson-like equation~\ref{sigma2de} quoted in the main text:
 \begin{equation}\label{sigma2sigma20}
  \begin{split}
   \sigma_{2}^{abc}(1,2,3)=&\iiint\sum_{def}\varepsilon^{-1,da}(1,4)\sigma_{2}^{\mathrm{KS},def}(4,5,6)\varepsilon^{-1,eb}(5,2)\varepsilon^{-1,fc}(6,3)d4d5d6\\&+\iiint\sum_{def}\sigma_{1}^{ad}(1,4)K_{\mathrm{xc},2}^{def}(4,5,6)\sigma_{1}^{eb}(5,2)\sigma_{1}^{fc}(6,3)d4d5d6.
  \end{split}
 \end{equation}
\end{widetext}
\section{KS OPTICAL RESPONSE TENSOR EXPRESSIONS UP TO SECOND ORDER}
\label{app:general}
In this appendix we provide the expressions of all optical KS response tensors up to second order within the formalism of Sipe and co-workers~\cite{PhysRevB.48.11705,PhysRevB.52.14636,PhysRevB.61.5337} (see Sec.~\ref{sec:optical}).
These expressions are valid for any combination of $\omega_{1}$ and $\omega_{2}$ and include metallic terms proportional to $k$-space derivatives of the occupation factors.
Here we merely quote the final expressions; for details on the derivation steps, we refer  the reader to Sec.~IV in Ref.~\cite{PhysRevB.61.5337} or to Appendix~A in the Supplemental Material of Ref.~\cite{PhysRevResearch.2.012017}.

At first order, the optical KS interband polarizability and intraband conductivity tensors are respectively expressed as
\begin{subequations}
 \begin{equation}\label{alphater10}
  \alpha_{\mathrm{ter},1}^{\mathrm{KS},ab}(\omega)=\frac{e^{2}}{\hbar V}\sum_{\mathbf{k}mn}f_{nm}\frac{r^{a}_{nm}r^{b}_{mn}}{\omega_{mn}-\tilde{\omega}},
 \end{equation}
 \begin{equation}\label{sigmatra10}
  \sigma_{\mathrm{tra},1}^{\mathrm{KS},ab}(\omega)=\frac{e^{2}}{\hbar V}\sum_{\mathbf{k}n}f_{n}\left(i\frac{\omega_{n;ab}}{\omega}-\epsilon_{cab}\Omega^{c}_{n}\right),
 \end{equation}
\end{subequations}
while at second order they are respectively expressed as
\begin{widetext}
 \begin{subequations}
  \begin{equation}\label{alphater20}
   \begin{split}
    \alpha_{\mathrm{ter},2}^{\mathrm{KS},abc}(\omega_{1},\omega_{2})=\frac{e^{3}}{2\hbar^{2}V}\bigg\{&\sum_{\mathbf{k}mnl}\frac{r^{a}_{nm}}{\omega_{mn}-\tilde{\omega}_{12}}\bigg[f_{nl}\bigg(\frac{r^{b}_{ln}r^{c}_{ml}}{\omega_{ln}-\tilde{\omega}_{1}}+\frac{r^{c}_{ln}r^{b}_{ml}}{\omega_{ln}-\tilde{\omega}_{2}}\bigg)-f_{lm}\bigg(\frac{r^{b}_{ml}r^{c}_{ln}}{\omega_{ml}-\tilde{\omega}_{1}}+\frac{r^{c}_{ml}r^{b}_{ln}}{\omega_{ml}-\tilde{\omega}_{2}}\bigg)\bigg]\\+i&\sum_{\mathbf{k}mn}\bigg(\frac{f_{nm}r^{b;c}_{mn}+f_{nm;c}r^{b}_{mn}}{\omega_{mn}-\tilde{\omega}_{1}}-\frac{f_{nm}r^{b}_{mn}\Lambda^{c}_{mn}}{\left(\omega_{mn}-\tilde{\omega}_{1}\right)^{2}}-\frac{f_{nm;b}r^{c}_{mn}}{\omega_{1}}\\&\quad\ +\frac{f_{nm}r^{c;b}_{mn}+f_{nm;b}r^{c}_{mn}}{\omega_{mn}-\tilde{\omega}_{2}}-\frac{f_{nm}r^{c}_{mn}\Lambda^{b}_{mn}}{\left(\omega_{mn}-\tilde{\omega}_{2}\right)^{2}}-\frac{f_{nm;c}r^{b}_{mn}}{\omega_{2}}\bigg)\bigg\}.
   \end{split}
  \end{equation}
  \begin{equation}\label{sigmatra20}
   \begin{split}
    \sigma_{\mathrm{tra},2}^{\mathrm{KS},abc}(\omega_{1},\omega_{2})=\frac{e^{3}}{2\hbar^{2}V}\bigg\{-&\sum_{\mathbf{k}mn}\bigg[\frac{f_{nm}\Lambda^{a}_{nm}}{\omega_{12}}\left(\frac{r^{c}_{nm}r^{b}_{mn}}{\omega_{mn}-\tilde{\omega}_{1}}+\frac{r^{b}_{nm}r^{c}_{mn}}{\omega_{mn}-\tilde{\omega}_{2}}\right)+f_{nm}\left(\frac{r^{c;a}_{nm}r^{b}_{mn}}{\omega_{mn}-\tilde{\omega}_{1}}+\frac{r^{b;a}_{nm}r^{c}_{mn}}{\omega_{mn}-\tilde{\omega}_{2}}\right)\bigg]\\+&\sum_{\mathbf{k}n}\left[i\left(\frac{f_{n;b}\epsilon_{dac}}{\omega_{1}}+\frac{f_{n;c}\epsilon_{dab}}{\omega_{2}}\right)\Omega^{d}_{n}-\frac{v^{a}_{n}f_{n;bc}}{\omega_{1}\omega_{2}}\right]\bigg\}.
   \end{split}
  \end{equation}
 \end{subequations}
\end{widetext}
All quantities appearing in the expressions above have been introduced in Sec.~\ref{sec:optical} except for $\omega_{n;ab}$ in Eq.~\ref{sigmatra10}, which stands for the inverse effective mass tensor.
As a remark, the metallic terms of the quadratic optical KS intraband polarizability tensor in Eq.~\ref{alphater20} are shown here for the first time to the best of our knowledge.
\section{TENSORIAL KERNELS}\label{app:xc}
Here we describe the calculation of the tensorial kernels in the optical limit.
Let us start by reviewing the Hartree contribution.
The Hartree potential is defined by
\begin{equation}
 V_{\mathrm{H}}(1)=\int v_{\mathrm{c}}(1,2)\rho(2)d2,
\end{equation}
where \mbox{$v_{\mathrm{c}}(1,2)=\delta(t_{1}-t_{2})/|\mathbf{r}_{1}-\mathbf{r}_{2}|$} is the static Coulomb scalar potential and $\rho(1)$ is the charge density.
With the aid of Maxwell's equation, \mbox{$\mathbf{E}(\mathbf{r},t)=-\boldsymbol{\nabla}V(\mathbf{r},t)$} and the continuity equation, \mbox{$\boldsymbol{\nabla}\cdot\mathbf{J}(\mathbf{r},t)=-\partial_{t}\rho(\mathbf{r},t)$}, the Hartree electric field in wavevector and frequency space is expressed as
\begin{equation}\label{ehfromvh}
 \mathbf{E}_{\mathrm{H}}(\mathbf{q}_{1},\omega)=\sum_{\mathbf{q}_{2}}
\overline{K}_{\mathrm{H}}(\mathbf{q}_{1},\mathbf{q}_{2},\omega)
 \cdot\mathbf{J}(\mathbf{q}_{2},\omega),
\end{equation}
with the kernel given by
\begin{equation}\label{Hkernel}
 K^{ab}_{\mathrm{H}}(\mathbf{q}_{1},\mathbf{q}_{2},\omega)={q}^{a}_{1}\frac{v_{\mathrm{c}}(\mathbf{q}_{1},\mathbf{q}_{2})}{i\omega}{q}^{b}_{2}={q}^{a}_{1}\frac{4\pi\delta_{\mathbf{q}_{1},\mathbf{q}_{2}}}{i\omega|\mathbf{q}_{1}||\mathbf{q}_{2}|}{q}^{b}_{2}.
\end{equation}
Above, $\mathbf{q}_{1}$ and $\mathbf{q}_{2}$ represent momenta and the Fourier transform of the Coulomb potential was used.
Applying  the $\mathbf{q}_{1},\mathbf{q}_{2}\to0$ optical limit in Eq.~\ref{Hkernel}, the tensorial Hartree kernel takes the usual form
\begin{equation}\label{Fh}
 {K}^{ab}_\mathrm{H}(\omega)=\delta_{ab}\dfrac{4\pi}{i\omega},
\end{equation}
which is a diagonal and isotropic tensor owing to the longitudinal and radial nature of the Coulomb force.

Coming now to the xc piece, its electric field up to linear order is written as
\begin{equation}\label{excfromvxc}
 {\mathbf{E}_{\mathrm{xc},1}}(\mathbf{q}_{1},\omega)=\sum_{\mathbf{q}_{2}}
\overline{K}_{\mathrm{xc},1}(\mathbf{q}_{1},\mathbf{q}_{2},\omega)
 \cdot\mathbf{J}(\mathbf{q}_{2},\omega).
\end{equation}
Assuming that the nonlocal long-range behaviour of excitonic effects completely dominates over all other terms in the optical limit~\cite{PhysRevLett.88.066404}, the xc contribution can be modelled by a Coulomb-like attractive interaction with LRC xc coefficients $\alpha^{a}_{\mathrm{LRC}}$.
In the wavevector and frequency domain, the corresponding tensorial xc kernel reads
\begin{equation}\label{xckernel}
 K^{ab}_{\mathrm{xc},1}(\mathbf{q}_{1},\mathbf{q}_{2},\omega)=-{q}^{a}_{1}\frac{\alpha^{a}_{\mathrm{LRC}}\delta_{ab}}{i\omega|\mathbf{q}_{1}||\mathbf{q}_{2}|}{q}^{b}_{2},
\end{equation}
which is a diagonal tensor owing to the longitudinal nature of Coulomb-like forces.
The tensorial nature of the xc kernel in TDCDFT was stressed in early works
by Vignale and co-workers~\cite{PhysRevLett.77.2037,PhysRevLett.79.4878}, 
as well as in later works making use of polarization functionals~\cite{PhysRevLett.115.137402,D0FD00073F}.
At variance with the Hartree contribution in Eq.~\ref{Hkernel}, the tensorial coefficients $\alpha^{a}_{\mathrm{LRC}}$ in Eq.~\ref{xckernel} allows a space-resolved anisotropic response of the xc electric field along the crystal axes.
By taking the optical limit in Eq.~\ref{xckernel}, we arrive at the
simplified expression used in our calculations,
\begin{equation}
 K^{ab}_{\mathrm{xc},1}(\omega)=-\frac{\alpha^{a}_{\mathrm{LRC}}\delta_{ab}}{i\omega}.
\end{equation}

The bootstrap method is a parameter-free approximation that was originally proposed for self-consistenty calculating the space-averaged isotropic scalar $\alpha$-coefficient in TDDFT~\cite{PhysRevLett.107.186401}.
We have adopted this method to compute $\alpha^{a}_{\mathrm{LRC}}$ by means of the expression
\begin{equation}\label{bootstrap}
 \alpha^{a}_{\mathrm{LRC}}=\epsilon^{-1,aa}_{\mathrm{M}}(0)\left[\overline{\alpha}^{\mathrm{KS}}_{1}(0)\right]^{-1,aa},
\end{equation}
which requires calculating the LRC xc coefficients independently for each of the three Cartesian directions.
As in the original bootstrap kernel, the calculation of the coefficients is done iteratively;
firstly, the microscopic optical MB conductivity is calculated by means of Eqs.~\ref{eq:eps_m} and \ref{eq:sigma1};
secondly, the macroscopic optical dielectric tensor by means of Eq.~\ref{eq:epsM};
and finally, the coefficients $\alpha^{a}_{\mathrm{LRC}}$ by means of Eq.~\ref{bootstrap}.
The iterative loop starts with the initial guess \mbox{$\alpha^{a}_{\mathrm{LRC}}=0$} and finishes when self-consistency is reached.
\section{THE OPTICAL MACROSCOPIC-MICROSCOPIC CONNECTION}
\label{appendix:micro-macro}
In this appendix we derive the relations that connect calculable response tensors at the microscopic scale with their measurable macroscopic counterparts in the optical limit.
This is largely based on the work of Del Sole and Fiorino for the first order~\cite{PhysRevB.29.4631}, and on the work of  Luppi and co-workers for the second order~\cite{PhysRevB.82.235201}.
\subsection{General definitions and useful relations}\label{J2P}
The response of a material to an applied external electric field can be mainly described in two ways.
On the one hand, the ability of a material to conduct an electric current is described by the electric conductivity, which relates the electric current-density vector to the electric field.
On the other hand, the ability of a material to electrically polarize is described by the electric susceptibility or polarizability, which relates the electric polarization-density vector to the electric field.

At the macroscopic scale (M), these relations are expressed in terms of the macroscopic total electric field $\mathbf{E}_{\mathrm{tot}}^{\mathrm{M}}(\mathbf{r},t)$, in such a way that the $j$-th order power series expansion of the macroscopic electric current- and polarization-density vectors are respectively defined as
\begin{subequations}
 \begin{equation}\label{JM}
  \begin{split}
   \mathbf{J}_{\mathrm{M},j}&(1)=\\&\int\!\!...\!\!\int^{1}_{0}\overline{\sigma}_{\mathrm{M},j}(1,...,j+1)\prod_{j}\mathbf{E}_{\mathrm{tot},\mathrm{M}}(j+1)dj+1
  \end{split}
 \end{equation}
 \begin{equation}\label{PM}
  \begin{split}
   \mathbf{P}_{\mathrm{M},j}&(1)=\\&\epsilon_{0}\int\!\!...\!\!\int^{1}_{0}\overline{\chi}_{j}(1,...,j+1)\prod_{j}\mathbf{E}_{\mathrm{tot},\mathrm{M}}(j+1)dj+1
  \end{split}
 \end{equation}
\end{subequations}
where \mbox{$\mathbf{J}_{\mathrm{M},j}(\mathbf{r},t)$} and \mbox{$\mathbf{P}_{\mathrm{M},j}(\mathbf{r},t)$} are the $j$-th order macroscopic electric current- and polarization-density vectors, respectively, and \mbox{$\overline{\sigma}_{\mathrm{M},j}(1,...,j+1)$} and \mbox{$\overline{\chi}_{j}(1,...,j+1)$} are the $j$-th order macroscopic conductivity and susceptibility tensors, respectively.
The complete macroscopic current- and polarization-density vectors are given by \mbox{$\mathbf{J}_{\mathrm{M}}(\mathbf{r},t)=\sum_{j}\mathbf{J}_{\mathrm{M},j}(\mathbf{r},t)$} and \mbox{$\mathbf{P}_{\mathrm{M}}(\mathbf{r},t)=\sum_{j}\mathbf{P}_{\mathrm{M},j}(\mathbf{r},t)$}, respectively.

In turn, at the microscopic scale the relations are expressed in terms of the microscopic external electric field $\mathbf{E}_{\mathrm{ext}}(\mathbf{r},t)$, in such a way that the $j$-th order power series expansion of the microscopic electric current- and polarization-density vectors are respectively defined as
\begin{subequations}
 \begin{equation}\label{Jm}
  \mathbf{J}_{j}(1)=\int\!\!...\!\!\int^{1}_{0}\overline{\sigma}_{j}(1,...,j+1)\prod_{j}\mathbf{E}_{\mathrm{ext}}(j+1)dj+1,
 \end{equation}
 \begin{equation}\label{Pm}
  \mathbf{P}_{j}(1)=\int\!\!...\!\!\int^{1}_{0}\overline{\alpha}_{j}(1,...,j+1)\prod_{j}\mathbf{E}_{\mathrm{ext}}(j+1)dj+1
 \end{equation}
\end{subequations}
where \mbox{$\mathbf{J}_{j}(\mathbf{r},t)$} and \mbox{$\mathbf{P}_{j}(\mathbf{r},t)$} are the $j$-th order microscopic electric current- and polarization-density vectors, respectively, and \mbox{$\overline{\sigma}_{j}(1,...,j+1)$} and \mbox{$\overline{\alpha}_{j}(1,...,j+1)$} are the $j$-th order microscopic conductivity and polarizability tensors, respectively.
The complete microscopic current- and polarization-density vectors are given by \mbox{$\mathbf{J}(\mathbf{r},t)=\sum_{j}\mathbf{J}_{j}(\mathbf{r},t)$} and \mbox{$\mathbf{P}(\mathbf{r},t)=\sum_{j}\mathbf{P}_{j}(\mathbf{r},t)$}, respectively.

In the absence of magnetization, and free charge and current densities, the  current- and polarization-density vectors are related by \mbox{$\mathbf{J}_{\mathrm{(M)},(j)}(\mathbf{r},t)=\partial_{t}\mathbf{P}_{\mathrm{(M)},(j)}(\mathbf{r},t)$}, both at the macroscopic and microscopic levels, as well as at any order of the power series expansion.
Using the latter relation and comparing Eq.~\ref{JM} and Eq.~\ref{PM}, we can derive the connections between the macroscopic conductivity and susceptibility up to second order.

In the reciprocal space and frequency domain, the connection at first order in the optical limit is given by
\begin{equation}\label{sigma12chie1}
 \overline{\sigma}_{\mathrm{M},1}(\omega)=-i\omega\epsilon_{0}\overline{\chi}_{1}(\omega),
\end{equation}
and at second order by
\begin{equation}\label{sigma22chie2}
 \overline{\sigma}_{\mathrm{M},2}(\omega_{1},\omega_{2})=-i\left(\omega_{1}+\omega_{2}\right)\epsilon_{0}\overline{\chi}_{2}(\omega_{1},\omega_{2}).
\end{equation}
In an analogous way, we can derive the connection between microscopic conductivity and polarizability up to second order, but this time comparing Eq.~\ref{Jm} and Eq.~\ref{Pm}.
At first order it is given by
\begin{equation}\label{sigma12alpha1}
 \overline{\sigma}_{1}(\omega)=-i\omega\overline{\alpha}_{1}(\omega),
\end{equation}
and at second order by
\begin{equation}\label{sigma22alpha2}
 \overline{\sigma}_{2}(\omega_{1},\omega_{2})=-i\left(\omega_{1}+\omega_{2}\right)\overline{\alpha}_{2}(\omega_{1},\omega_{2}).
\end{equation}
\subsection{Macroscopic optical susceptibility}
Our main goal is to express macroscopic response tensors as a function of their respective microscopic counterpart.
To this end, the simplest option is to switch to the KS electronic system, where the observables in Eqs.~\ref{Jm} and \ref{Pm} are defined in terms of the microscopic total electric field $\mathbf{E}_{\mathrm{tot}}(\mathbf{r},t)$ as in Eq.~\ref{jneqsigmaksprodnetot} for the current, and then take a macroscopic spatial average of the microscopic quantities.
In the so-called long-wavelength limit, where the real-space variation of the total electric field over distances of the order of the lattice parameter is neglected and therefore the total electric field is \textit{per se} of macroscopic character, the macroscopic spatial average of microscopic quantities is straightforward; it is sufficient to retain the $\mathbf{G}=0$ reciprocal lattice vector~\cite{Ehrenreich}.
Furthermore, the averaging is even more direct in the optical limit, since microscopic quantities are calculated assuming ideally a non-variational character in space.
Therefore, under this point of view, one can state that the macroscopic optical conductivity is equal to its microscopic KS counterpart at any order, \textit{i.e.}~\mbox{$\sigma_{\mathrm{M},j}(1,...,j+1)=\sigma^{\mathrm{KS}}_{j}(1,...,j+1)$}.

Nevertheless, the previous approach does not account for many-body effects in the response, since those are assumed to be already included in the total electric field.
In order to overcome this limitation, one can obtain an expression of the external electric field as a function of the total electric field at the microscopic level by using Maxwell's equations and related constitutive relations.
Then, the resulting expression is used to define microscopic observables in Eqs.~\ref{Jm} and \ref{Pm} in terms of the total electric field, whose macroscopic spatial averages give access to the formulation of macroscopic response tensors including many-body effects.
Following Ref.~\cite{PhysRevB.82.235201}, in the reciprocal space and frequency domain, the longitudinal-longitudinal $(LL)$ component of the linear macroscopic susceptibility tensor is given by~\cite{PhysRevB.29.4631}
\begin{equation}
 \chi^{LL}_{1}(\mathbf{q},\omega)=4\pi\alpha^{LL}_{1}(\mathbf{q},\omega)\epsilon^{LL}_{\mathrm{M}}(\mathbf{q},\omega),
\end{equation}
where \mbox{$\alpha^{LL}_{1}(\mathbf{q},\omega)\equiv{\alpha_{1}}^{LL}_{\mathbf{G}\mathbf{G'}}(\mathbf{q},\omega)\delta_{\mathbf{G},0}\delta_{\mathbf{G'},0}$} is the $LL$ component of the macroscopic spatial averaged microscopic MB polarizability tensor at first order, and \mbox{$\epsilon^{LL}_{\mathrm{M}}(\mathbf{q},\omega)=[1-4\pi\alpha^{LL}_{1}(\mathbf{q},\omega)]^{-1}$} is the $LL$ component of the macroscopic dielectric tensor.
In an analogous way, the longitudinal-longitudinal-longitudinal $(LLL)$ component of the quadratic macroscopic susceptibility tensor is expressed as
\begin{widetext}
 \begin{equation}
  \chi^{L_{12}L_{1}L_{2}}_{2}(\mathbf{q}_{1},\mathbf{q}_{2},\omega_{1},\omega_{2})=4\pi\epsilon^{L_{12}L_{12}}_{\mathrm{M}}(\mathbf{q}_{12},\omega_{12})\alpha_{2}^{L_{12}L_{1}L_{2}}(\mathbf{q}_{1},\mathbf{q}_{2},\omega_{1},\omega_{2})\epsilon^{L_{1}L_{1}}_{\mathrm{M}}(\mathbf{q}_{1},\omega_{1})\epsilon^{L_{2}L_{2}}_{\mathrm{M}}(\mathbf{q}_{2},\omega_{2}),
 \end{equation}
where $L_{1}$, $L_{2}$ and $L_{12}$ stand for the longitudinal component along the directions $\mathbf{q}_{1}$, $\mathbf{q}_{2}$ and $\mathbf{q}_{12}\equiv\mathbf{q}_{1}+\mathbf{q}_{2}$, respectively, and \mbox{$\alpha^{L_{12}L_{1}L_{2}}_{2}(\mathbf{q}_{1},\mathbf{q}_{2},\omega_{1},\omega_{2})\equiv{\alpha_{2}}^{L_{12}L_{1}L_{2}}_{\mathbf{G}_{12}\mathbf{G}_{1}\mathbf{G}_{2}}(\mathbf{q}_{1},\mathbf{q}_{2},\omega_{1},\omega_{2})\delta_{\mathbf{G}_{12},0}\delta_{\mathbf{G}_{1},0}\delta_{\mathbf{G}_{2},0}$} is the $LLL$ component of the spatially averaged microscopic MB polarizability tensor at second order.
\end{widetext}

The adopted framework is valid for any $\mathbf{q}$ and describes longitudinal responses to longitudinal perturbations~\cite{PhysRevB.29.4631}.
In the optical limit $(\mathbf{q}\to0)$, one can always find three principal axes for any crystal symmetry in which the macroscopic dielectric tensor is diagonal~\cite{Wooten}.
In this reference frame a longitudinal perturbation induces a longitudinal response, hence any optical property of the crystal can be deduced from a longitudinal calculation~\cite{Botti2012}.
Therefore, in the principal frame the linear macroscopic optical susceptibility tensor is expressed as
\begin{equation}\label{chie1aa}
 \chi^{aa}_{1}(\omega)=4\pi\alpha^{aa}_{1}(\omega)\epsilon^{aa}_{\mathrm{M}}(\omega),
\end{equation}
and the quadratic macroscopic optical susceptibility tensor as
\begin{equation}\label{chie2aaa}
 \begin{split}
  \chi^{abc}_{2}(\omega_{1},\omega_{2})=4\pi\epsilon_{\mathrm{M}}^{aa}(\omega_{12})&\alpha_{2}^{abc}(\omega_{1},\omega_{2})\\&\times\epsilon_{\mathrm{M}}^{bb}(\omega_{1})\epsilon_{\mathrm{M}}^{cc}(\omega_{2}),
 \end{split}
\end{equation}
where $a$, $b$ and $c$ are principal axis components of the crystal.
Note that for any crystal with a symmetry greater or equal to the orthorhombic symmetry, $a$, $b$ and $c$ coincide with the Cartesian coordinates~\cite{boyd2008nonlinear}.
\subsection{Macroscopic optical conductivity}
The derivation of the optical macroscopic-microscopic connection 
in the previous section has been given
in terms of the macroscopic susceptibility and the microscopic polarizability.
Nevertheless, one can also express this 
connection in terms of the conductivity by means of the identities provided in Sec.~\ref{J2P}.
In particular, inserting Eqs.~\ref{sigma12chie1} and \ref{sigma12alpha1} into Eq.~\ref{chie1aa} one obtains the linear macroscopic optical conductivity, 
\begin{equation}
 \sigma_{\mathrm{M},1}^{aa}(\omega)=\sigma^{aa}_{1}(\omega)\epsilon^{aa}_{\mathrm{M}}(\omega),
\end{equation}
while inserting Eqs.~\ref{sigma22chie2} and \ref{sigma22alpha2} into Eq.~\ref{chie2aaa} yields the expression for 
the quadratic macroscopic optical conductivity,
\begin{equation}\label{sigma2M}
 \begin{split}
  \sigma_{\mathrm{M},2}^{abc}(\omega_{1},\omega_{2})=\epsilon^{aa}_{\mathrm{M}}(\omega_{12})&\sigma^{abc}_{2}(\omega_{1},\omega_{2})\\&\times\epsilon^{bb}_{\mathrm{M}}(\omega_{1})\epsilon^{cc}_{\mathrm{M}}(\omega_{2}).
 \end{split}
\end{equation}
\bibliography{article}
\end{document}